\documentclass[aps,pre,amsmath,amsfonts,amssymb,nofootinbib]{revtex4}

\usepackage{graphicx}
\usepackage{epsfig}

\def\s{\sigma}
\def\us{\underline{\sigma}}
\def\t{\tau}
\def\ut{\underline{\tau}}

\def\eqd{\overset{\rm d}{=}}
\def\tod{\overset{\rm d}{\to}}
\def\E{\mathbb{E}}
\def\I{\mathbb{I}}
\def\T{\mathbb{T}}
\def\hT{\widehat{\mathbb{T}}}

\def\M{{\cal M}}
\def\P{{\cal P}}           

\def\X{{\cal X}}
\def\S{{\cal S}}

\def\oeta{\overline{\eta}}

\def\tP{\widetilde{P}}

\def\tf{\widetilde{f}}
\def\tg{\widetilde{g}}
\def\hq{\widehat{q}}
\def\hQ{\widehat{Q}}
\def\hR{\widehat{R}}
\def\hP{\widehat{P}}
\def\hPP{\widehat{\cal P}}
\def\hphi{\widehat{\phi}}
\def\tphi{\widetilde{\phi}}

\begin{document}
\date{\today}
\title{On the freezing of variables in random constraint satisfaction problems}
\author{Guilhem Semerjian}
\affiliation{LPTENS, Unit\'e Mixte de Recherche (UMR 8549) du CNRS et de l'ENS
associ\'ee \`a l'universit\'e Pierre et Marie Curie, 24 Rue Lhomond, 75231 
Paris Cedex 05, France}

\begin{abstract}
The set of solutions of random constraint satisfaction problems
(zero energy groundstates of mean-field diluted spin glasses)
undergoes several structural phase transitions as the amount
of constraints is increased. This set first breaks down into a large
number of well separated clusters. At the freezing transition, which is
in general distinct from the clustering one, some variables
(spins) take the same value in all solutions of a given cluster. In this 
paper we study the critical behavior around the freezing transition, 
which appears in the unfrozen phase as the divergence of the sizes of
the rearrangements induced in response to
the modification of a variable. The formalism is developed on generic
constraint satisfaction problems and applied
in particular to the random satisfiability of boolean formulas and to the 
coloring of random graphs. The computation is first performed in random tree
ensembles, for which we underline a connection with percolation models and 
with the reconstruction problem of information theory.
The validity of these results for the original random ensembles
is then discussed in the framework of the cavity method.

\end{abstract}

\maketitle

\section{Introduction}

The theory of computational complexity~\cite{NP} establishes a classification
of constraint satisfaction problems (CSP) according to their difficulty in the 
worst case. For concreteness let us introduce the three problems we shall 
use as running examples in the paper:
\begin{itemize}
\item[$\bullet$] $k$-XORSAT. 
Find a vector $\vec{x}$ of boolean variables satisfying the linear equations 
$A \vec{x}=\vec{b} \ ({\rm mod} \ 2)$, where each row of the $0/1$ matrix $A$ 
contains exactly $k$ non-null elements, and $\vec{b}$ is a given boolean
vector.

\item[$\bullet$] $q$-coloring ($q$-COL). 
Given a graph, assign one of $q$ colors 
to each of its vertices, without giving the same color to the two extremities 
of an edge.

\item[$\bullet$] $k$-satisfiability ($k$-SAT). 
Find a solution of a boolean formula 
made of the conjunction (logical AND) of clauses, each made of the disjunction 
(logical OR) of $k$ literals (a variable or its logical negation).
\end{itemize}
Each of these problems admits several variants. In the decision version
one has to assert the existence or not of a solution,
for instance a proper coloring of a given graph. More elaborate questions
are the estimation of the number of such solutions, or, in the absence of
solution, the discovery of optimal configurations, for instance colorings
minimizing the number of monochromatic edges. The decision variant of the 
three examples stated above fall into two distinct complexity classes: 
$k$-XORSAT is in the P class, while the two others are NP-complete for 
$k,q \ge 3$ (see~\cite{CSP} for a classification of generic boolean CSPs).
This means that the existence of a solution of the XORSAT problem can
be decided in a time growing polynomially with the number of variables,
for any instance of the problem; one can indeed use the Gaussian elimination 
algorithm. On the contrary no fast algorithm able of solving every coloring or
satisfiability problem is known, and the existence of such a polynomial
time algorithm is considered as highly improbable.

This notion of computational complexity, being based on worst-case 
considerations, could overlook the possibility that ``most'' of the
instances of an NP problem are in fact easy and that the difficult cases are
very rare. Random ensembles of problems have thus been 
introduced in order to give a quantitative content to this notion of typical 
instances; a
property of a problem will be considered as typical if its probability
(with respect to the random choice of the instance) goes to one in the limit 
of large problem sizes. Most random ensembles depend on an external parameter
that can be varied continuously. In the coloring problem one can for
instance consider the traditional Erd\"os-R\'enyi random 
graphs~\cite{random_graphs} which are
parameterized by their mean connectivity $c$. For (XOR)SAT instances this
role is played by the ratio $\alpha$ of the number of constraints (clauses
for SAT or rows in the matrix for XORSAT) to the number of variables.
A remarkable threshold phenomenon, first observed 
numerically~\cite{randomsat}, occurs when this
parameter is varied: when a particular value $c_{\rm s},\alpha_{\rm s}$ is
crossed from below, the instances go from typically satisfiable to typically
unsatisfiable. This statement has been rigorously proven for 
XORSAT~\cite{xor1,xor2} and for 2-SAT~\cite{2sat}, in the other cases it is 
only a largely accepted conjecture, with sharpness condition on the width
of the transition window~\cite{Friedgut} and bounds on its possible 
location~\cite{lbounds,ubounds}.

Threshold phenomena are largely studied in statistical mechanics under the
name of phase transitions. There is moreover a natural analogy between 
optimization problems and statistical mechanics; if one defines the energy
as the number of violated constraints, for instance the number of 
monochromatic edges, the optimal configurations of a
problem coincide with the groundstates of the associated physical system, 
an antiferromagnetic Potts model in the coloring case. This analogy
triggered a large amount of research, relying on methods of statistical
mechanics of disordered systems originally devised for the study of 
mean-field spin-glasses~\cite{beyond}. 
Early examples of this approach for the satisfiability
and coloring problems can be found in~\cite{sat_first,col_first}.

One of the most interesting outcomes of this line of 
research~\cite{BiMoWe,SP_science} 
has been the
suggestion that other structural threshold phenomenon take place before
the satisfiability one\footnote{It was of course already known that the 
algorithms rigorously studied to derive lower bounds on the satisfiability
threshold work only upto to values of $\alpha$ smaller than 
$\alpha_{\rm s}$~\cite{lbounds}. These values
are however largely algorithm-dependent and not directly related to a
change of structure in the configuration space.}.
The set of solutions
of a random CSP, viewed as a subset of the whole configuration space,
is smooth at low values of the constraint ratio but becomes fragmented into 
clusters of solutions for intermediate values of the control parameter,
$\alpha \in [\alpha_{\rm d},\alpha_{\rm s}]$. This clustering transition has 
been rigorously demonstrated in the XORSAT case~\cite{xor1,xor2}, 
for which it has a simple
geometric interpretation. $\alpha_{\rm d}$ is indeed the threshold for
the percolation of the 2-core of the hypergraph underlying the CSP;
between $\alpha_{\rm d}$ and $\alpha_{\rm s}$ there is typically a
finite fraction of the variables and constraints in a peculiar sub-formula
known as the backbone. Every solution of the backbone gives birth to a cluster
of the complete formula. The variables of the backbone are said to be 
frozen in a given cluster, i.e. they take the same value in all the
solutions belonging to a cluster; this is merely a consequence of the 
definition of a cluster in this case.

Establishing a precise and generic definition of the clusters is not an easy
task, not to speak about proving tight rigorous results on their existence or
properties (for recent results in this direction 
see~\cite{Maneva,MeMoZe,AcRi,Papadimitriou}). 
Even at the heuristic
level, it was recently argued~\cite{letter,long_col,long_sat} 
that the computation of $\alpha_{\rm d}$
for random satisfiability (or $c_{\rm d}$ for coloring) by previous statistical
mechanics studies~\cite{MeMeZe,thresh_col} was incorrect. 
Roughly speaking, in these two models,
the sizes of the clusters can have large fluctuations~\cite{PaMeRi} 
that must be taken into
consideration. In~\cite{letter} the existence of yet another threshold 
(for $k,q \ge 4$) $\alpha_{\rm c} \in [\alpha_{\rm d},\alpha_{\rm s}]$
was also pointed out; this condensation threshold separates two clustered 
regimes, one where the relevant clusters are exponentially numerous (for 
smaller values of $\alpha$) and the other where there is only a 
sub-exponential number of them.

The clustering transition of XORSAT, because of its geometric interpretation,
is certainly a good example on which developing one's intuition of
the clustering phenomenon. There are however at least two aspects in which
XORSAT departs from other CSP and where the intuitive picture must be taken 
with a grain of salt. The first is that the clusters of XORSAT all have
the same size, because of the linear algebra structure of its set of solutions.
For this reason the condensation phenomenon is not present in XORSAT. The
second point is that clusters of XORSAT have frozen variables, by
definition. There is however no obvious reason that this should be true
for any CSP. On the contrary we shall argue in this paper that in general
frozen variables appear at another value $\alpha_{\rm f}$ of the control
parameter, with generically 
$\alpha_{\rm f} \in [\alpha_{\rm d}, \alpha_{\rm s}]$. This was one of the
results of~\cite{long_col,long_sat}, here we shall develop this point and 
quantify the precursors of the transition before $\alpha_{\rm f}$. 
For this we build upon the study of XORSAT presented in~\cite{MoSe} and
extend it to generic CSPs, in particular satisfiability and coloring.
The central notion studied here is the one of rearrangement
(to some extent related to the long-range frustration of~\cite{Zhou}): 
given an initial
solution of a CSP and a variable $i$ that one would like to modify, a 
rearrangement is a path in configuration space that starts from the initial
solution and leads to another solution where the value of the $i$'th variable
is changed with respect to the initial one. The minimal length of such
a path is a measure of how constrained was the variable $i$ in the initial 
configuration. In intuitive terms this length diverges with the system size
when the variable was frozen in the initial cluster.

The paper is organized as follows. In Sec.~\ref{sec_def} we introduce a
generic class of CSPs and precise the definition of
the rearrangements. Sections~\ref{sec_tree} and 
\ref{sec_results} are devoted to modified (tree) random ensembles in which
the approach is essentially rigorous; the former
presents detailed computations in a rather generic setting and its application
to the three selected examples, while the latter presents the numerical results
and discuss the generic phenomenology at the approach of the freezing 
transition in the tree ensembles, with some more technical details deferred
to App.~\ref{sec_app}. The computation is reconsidered in the perspective of
the reconstruction problem in Sec.~\ref{sec_reconstruction}.
The applicability of these results to the
original ensembles is discussed in Sec.~\ref{sec_rgraph}, through a precise
statement of the hypotheses of the cavity method. Conclusions and perspectives
for future work are presented in Sec.~\ref{sec_conclu}.

\section{Definitions}
\label{sec_def}

We introduce here some notations and definitions for a class of problems
that encompasses the three examples we shall treat in more details.
The degrees of freedom of the CSP will be $N$ variables $\s_i$ taking
values in a discrete alphabet $\X$; global configurations are denoted 
$\us=(\s_1,\dots,\s_N)$.
An instance (or formula) $F$ of the CSP is a set of $M$ constraints between
the variables $\s_i$. The $a$'th constraint is defined by a function
$\psi_a(\us_a) \to \{0,1\}$, which depends on the configuration of a subset
of the variables $\us_a$ and is equal to 1 if the constraint is satisfied,
0 otherwise. The set $\S_F \subset \X^N $ of solutions of $F$ is composed
of the configurations satisfying simultaneously all the constraints. It can
thus be formally defined as $\S_F =\{\us | \psi_F(\us) =1  \}$, where the
indicator function $\psi_F$ is
\begin{equation}
\psi_F( \us) = \prod_{a=1}^M \psi_a(\us_a) \ .
\end{equation}
When the formula admits a positive number of solutions, call it
$Z_F$, the uniform measure over the solutions is denoted
$\mu_F(\us) = \psi_F(\us)/Z_F$.

\begin{figure}
\includegraphics[width=6cm]{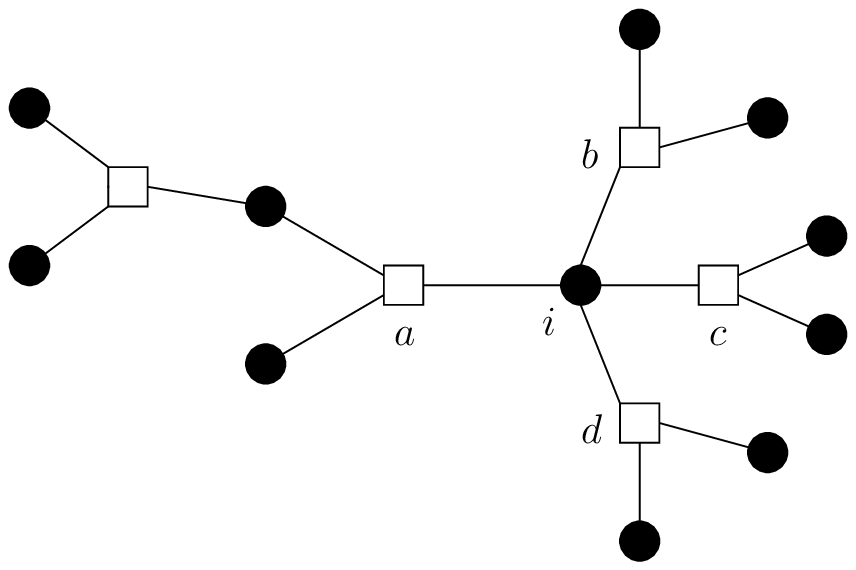}
\caption{An example of factor graph. The neighborhoods are for instance 
$\partial i = \{a,b,c,d\}$ and $\partial i \setminus a = \{b,c,d\}$}
\label{fig_factor}
\end{figure}

Factor graphs~\cite{fgraph} provide an useful representation of a CSP.
These graphs (see Fig.~\ref{fig_factor} for an example) have two kind of 
nodes. Variable nodes (filled circles on the figure) are associated to
the degrees of freedom $\s_i$, while constraint nodes
(empty squares) represent the clauses $\psi_a$. An edge between constraint
$a$ and variable $i$ is drawn whenever $\psi_a$ depends on $\s_i$. 
The neighborhood $\partial a$ of a constraint node is the set of variable 
nodes that appear in $\us_a$. Conversely $\partial i$ is the set of 
constraints that depend on $\s_i$. We shall conventionally use the indices
$i,j,\dots$ for the variable nodes, $a,b,\dots$ for the constraints,
and denote $\setminus$ the subtraction from a set. Two variable nodes
are called adjacent if they appear in a common constraint. The graph distance
between two variable nodes $i$ and $j$ is the number of constraint nodes
encountered on a shortest path linking $i$ and $j$ (formally infinite if the
two variables are not in the same connected component of the graph).

The three illustrative examples presented above admits a simple representation
in this formalism:
\begin{itemize}
\item[$\bullet$] $k$-XORSAT. The degrees of freedom of this CSP are boolean
variables that we shall represent, following the physics conventions, by
Ising spins, $\X=\{-1,+1\}$. 
Each constraint involves a subset of $k$ variables,
$\us_a=(\s_{i_a^1},\dots,\s_{i_a^k})$, and reads 
$\psi_a(\us_a) = \I( \s_{i_a^1} \dots \s_{i_a^k}  = J_a)$, where here and in
the following $\I(\cdot)$
denotes the indicator function of an event and $J_a \in\{-1,+1\} $ is a given
constant. This is equivalent to the definition given in 
the introduction: defining $x_i,b_a \in \{0,1 \}$ such that 
$\s_i = (-1)^{x_i}$ and $J_a=(-1)^{b_a}$, the constraint imposed by $\psi_a$
reads $x_{i_a^1} + \dots + x_{i_a^k} = b_a \pmod 2$, which is nothing but 
the $a$'th row of the matrix equation $A \vec{x} = \vec{b}$. The addition
modulo 2 of Boolean variables can also be read as the binary exclusive OR 
operation, hence the name XORSAT used for this problem.

\item[$\bullet$] $q$-COL. Here $\X=\{1,\dots,q\}$ is the set of allowed
colors on the $N$ vertices of a graph. Each edge $a$ connecting the vertices
$i$ and $j$ prevents them from being of the same color: 
$\psi_a(\s_i,\s_j) = \I(\s_i \neq \s_j)$.

\item[$\bullet$] $k$-SAT. As in the XORSAT problem one deals with Ising 
represented boolean variables, but in each clause the XOR operation between
variables is replaced by an OR between literals (i.e. a variable or its 
negation). In other words a constraint $a$ is unsatisfied only when all 
literals evaluate to false, or in Ising terms when all spins $\s_i$
involved in the constraint take their wrong value that we denote $J_a^i$:
 $\psi_a(\us_a) =  1-\I ( \s_i = J_a^i \ \ \forall i \in \partial a )$.

\end{itemize}

The random ensembles of CSPs instances we shall use are defined as follows:
\begin{itemize}
\item[$\bullet$] $k$-XORSAT. For each of the $M$ clauses $a$ a $k$-uplet of 
distinct 
variable indices $(i_a^1,\dots,i_a^k)$ is chosen uniformly at random among the
$\binom{N}{k}$ possible ones, and the constant
$J_a$ is taken to be $\pm 1$ with probability one-half.

\item[$\bullet$] $q$-coloring. A set of $M$ among the $\binom{N}{2}$ possible
edges $a=\{i,j\}$ is chosen uniformly at random.

\item[$\bullet$] $k$-SAT. The variables $i_a^j$ are chosen as in the XORSAT 
ensemble, and the $J_a^i$ are independently taken to be $\pm 1$ with 
equal probability. 
\end{itemize}
For the coloring problem this construction is the classical Erd\"os-R\'enyi
random graph $G(N,M)$, the two other cases are its random hypergraph 
generalization. We are interested in the thermodynamic limit of large
instances where $N$ and $M$ both diverge with a fixed ratio 
$\alpha=M/N$~\footnote{In this limit the quantities studied in this paper
are not affected by some variations around these models. For instance in the
coloring case $G(N,M)$ can be replaced by the ensemble $G(N,p)$ where each edge
is present independently with probability $p=2\alpha/N$, such that
the average number of edges is close to $M$. The choice of the (hyper)edges
with or without replacement is also irrelevant.}. Random (hyper)graphs
have many interesting properties in this limit~\cite{random_graphs}.
For instance the degree of a variable node of the factor graph converges to 
a Poisson law of average $\alpha k$ for the XORSAT and SAT cases, 
and $2 \alpha$ for the coloring ensemble. For clarity in the latter case 
we shall use the notation $c=2\alpha$ for the average connectivity. 
Moreover, picking at random one variable node $i$ and isolating the subgraph 
induced by the variable nodes at a graph distance smaller than a given constant
$L$ yields, with a probability going to one in the thermodynamic limit, a
(random) tree. This tree can be described by a Galton-Watson branching 
process: 
the root $i$ belongs to $l$ constraints, where $l$ is a Poisson random variable
of parameter $\alpha k$ ($c$ in the coloring case). The variable nodes
adjacent to $i$ give themselves birth to new constraints, in numbers which
are independently Poisson distributed with the same parameter. This
reproduction process is iterated on $L$ generations, until the variable nodes
at graph distance $L$ from the initial root $i$ have been generated.

We now define the main object of our study. First recall the well-known
definition of the Hamming distance between two configurations,
$d(\us,\ut) = \sum_{i=1}^N \I(\s_i \neq \t_i )$. Consider an initial solution 
of the formula, $\us \in \S_F$, and imagine one wants to modify the value
of the variable $i$. A rearranged solution is a new configuration 
$\ut \in \S_F $ such that $\t_i \neq \s_i$. The minimal size of a 
rearrangement (m.s.r.) for variable $i$ starting from $\us \in \S_F$ is
defined as
\begin{equation}
n_i(\us,F) = \min_{\ut} \{ d(\us,\ut) | \ut \in \S_F , \t_i \neq \s_i \} \ ,
\end{equation}
and measures how costly (in terms of Hamming distance) it is to perturb
the solution at variable $i$~\footnote{if $\s_i$ takes the same value in every
solution we formally define $n_i=N+1$.}. 
It can also be viewed as the minimal length
of a path in configuration space, modifying one variable at a time, between
$\us$ and another solution with a different value of variable $i$, thus
providing a quantification of how much constrained was initially this variable.
We shall also speak of the support of a rearrangement as the set of variables 
which differ in the initial and final configurations, the size of the
rearrangement being the cardinality of its support.

In general the m.s.r. will depend on the starting configuration, we thus 
define its distribution with respect to an uniform choice of $\us$ 
(in abbreviation m.s.r.d.),
\begin{equation}
q_n^{(i,F)} = \sum_{\us} \mu_F(\us) \ \delta_{n,n_i(\us,F)} \ .
\end{equation}
There should be no possibility of confusion between the distribution $q_n$ 
and the number $q$ of allowed colors in the $q$-COL problem.
When dealing with random CSPs we shall study the average of this 
distribution,
\begin{equation}
q_n = \E \ q_n^{(i,F)} \ ,
\label{eq_def_qn_rg}
\end{equation}
where the expectation is taken with respect to the instance ensemble (in the
cases considered here all variable nodes are equivalent on average). 
Its behavior in the thermodynamic limit will drastically
change with the connectivity parameter $\alpha$ (or $c$ for the coloring).
We shall indeed define the threshold $\alpha_{\rm f}$ ($c_{\rm f}$) as the 
value above which a finite fraction of the distribution $q_n$ is supported on
sizes $n$ that diverge with the number of variables. In pictorial terms
clusters acquire frozen variables at this point, their rearrangements must
be of diverging size and thus lead to a final solution outside the initial
cluster.

The computation of the average m.s.r.d. will be first undertaken in a random
tree ensemble, mimicking the tree neighborhoods of the
random graphs. The threshold for the freezing transition in these tree 
instances will be computed, along with a set of exponents characterizing
the behavior of the average m.s.r.d. when the transition is approached from
the unfrozen phase. For clarity we shall denote $\alpha_{\rm p}$ instead of
$\alpha_{\rm f}$ the thresholds in the tree ensembles.
We shall then argue in Sec.~\ref{sec_rgraph}, on the basis
of the non-rigorous cavity method, that for some values of $\alpha$ and $k$ 
the properties of the random graphs instances are correctly described
by the computations in the tree ensemble. In particular for large enough
values of $k$ we shall conjecture that $\alpha_{\rm p}=\alpha_{\rm f}$.
We will also explain how the 
computation has to be amended to handle the more elaborated version of the 
cavity method (with replica-symmetry breaking), and what are the expectedly
universal characteristics of the critical behavior at the freezing transition.

\section{Minimal size rearrangements in random tree ensembles}
\label{sec_tree}

\begin{figure}
\includegraphics[width=8cm]{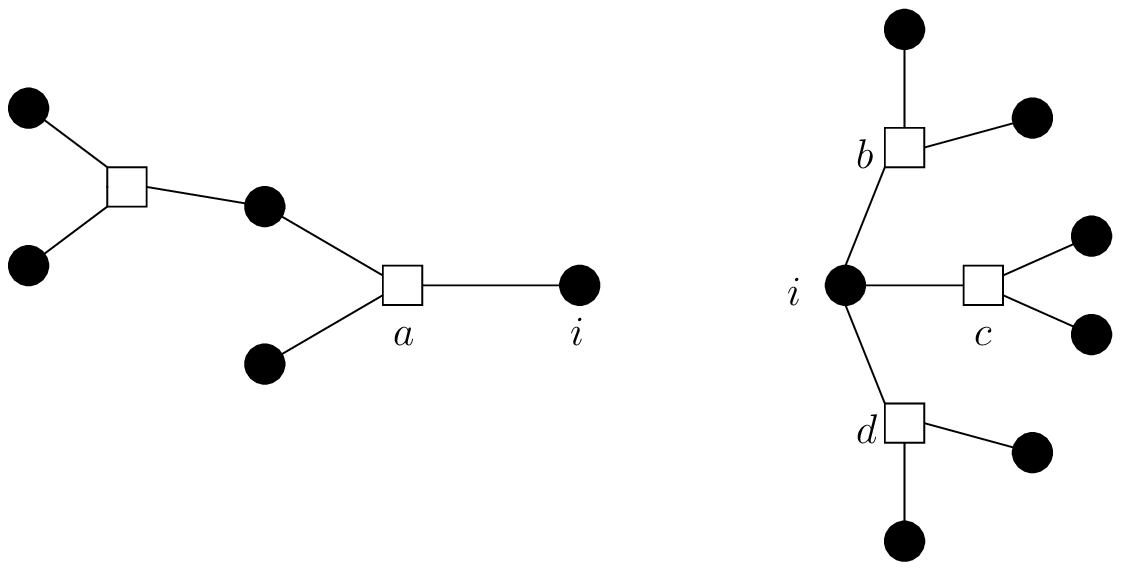}
\caption{The cavity graphs $F_{a \to i}$ and $F_{i \to a}$ obtained from
the example of Fig.~\ref{fig_factor}.}
\label{fig_cavity}
\end{figure}

In this and the next Section all the instances of CSP encountered have an 
underlying factor graph which is a finite tree. 
Given such a formula $F$ (or equivalently its
factor graph) and an edge $i-a$ between a variable node $i$ and
an adjacent constraint node $a$, we define two sub formulas
(cavity graphs)
$F_{i \to a}$ and $F_{a \to i}$. $F_{i \to a}$ is obtained from $F$ by 
deleting the branch of the formula rooted at $i$ starting with constraint $a$.
Conversely $F_{a \to i}$ is obtained by keeping only this branch 
(see Fig.~\ref{fig_cavity}). We also decompose the configuration $\us$ 
as $(\us_{a \to i},\s_i,\us_{i \to a})$, where $\us_{a \to i}$
(resp. $\us_{i \to a}$) is the configurations of the variable nodes in 
$F_{a \to i}$ (resp. $F_{i \to a}$) distinct from $i$. The
notation $\us_{\setminus i}$ will be used for the configuration of all
variables except $i$.
The computation, based on the natural recursive structure of trees, will 
be performed in three steps: we shall first see how to obtain $n_i(\us,F)$, 
then its distribution with respect to $\us$, $q_n^{(i,F)}$, which shall
finally be averaged over a random tree ensemble. For notational simplicity
$F$ will often be kept implicit. This approach is
presented in a general setting before the three specific cases of XORSAT, COL
and SAT are treated.

\subsection{General case}

\subsubsection{Given tree, given $\us$}

The computation of the m.s.r. $n_i$ on a tree factor graph can be performed in 
a recursive way. One has to determine, for each value of $\t_i\neq \s_i$, the
cost, in terms of Hamming distance, of the modification $\s_i \to \t_i$.
This can be done by computing separately these costs in the factor graphs 
$F_{a \to i}$ for all the constraint nodes $a$ around $i$ and then patching
together the rearrangements of the sub-formulae.
Rearranging a factor graph $F_{a \to i}$ amounts to looking for a configuration
of the variables $j \in \partial a \setminus i$ which satisfies the interaction
$a$ and which provokes a minimal propagation of the rearrangement in the 
branches $F_{j \to a}$.

To formalize this reasoning we introduce a $q$-component vectorial notation,
$\vec{n}$, where the rows of the vectors are indexed by a spin value in $\X$, 
and we shall denote $[\vec{n}]_\t$ the $\t^{\rm th}$ component of $\vec{n}$.
We define $\vec{n}_i(\us)$ as the m.s.r. for $i$ starting from the
initial configuration $\us$, and with the final value $\t_i$ encoded in the
row of the vector:
\begin{equation}
[\vec{n}_i(\us) ]_{ \t_i } = \min_{\ut_{\setminus i}} 
\{ d(\us,\ut=(\t_i,\ut_{\setminus i})) | \ut \in S_F \} \ .
\end{equation}
The original quantity $n_i(\us)$ is obtained from this more detailed one
as $n_i(\us) = \underset{\t_i \neq \s_i }{\min} [\vec{n}_i(\us) ]_{ \t_i }$. 
The
recursive computation of $\vec{n}_i$ is performed in terms of
vectorial messages on the directed edges of the factor graph, 
$\vec{n}_{i \to a}$ and $\vec{n}_{a \to i}$. The former, 
$\vec{n}_{i \to a}(\s_i,\us_{i \to a})$ is defined exactly as $\vec{n}_{i}$
with the cavity graph $F_{i \to a}$ replacing the original formula $F$.
The latter reads
\begin{equation}
[\vec{n}_{a \to i}(\us_{a \to i})]_{\t_i} = \min_{\ut_{a \to i}} \{
d(\us_{a \to i},\ut_{a \to i}) | 
(\t_i,\ut_{a \to i}) \in \S_{F_{a \to i}} \} \ .
\end{equation}
Note that here one does not count the cost of flipping the root variable,
which avoids overcounting when gluing together the cavity graphs.
A moment of thought reveals that these messages obey the following recursive
equations:
\begin{equation}
\vec{n}_{a \to i}(\us_{a \to i}) = \tf(\{ 
\vec{n}_{j \to a}(\s_j,\us_{j \to a}) \}_{j \in \partial a \setminus i}) 
\ , \qquad \qquad
\vec{n}_{i \to a}(\s_i,\us_{i \to a}) = \tg_{\s_i}(\{ 
\vec{n}_{b \to i}(\us_{b \to i}) \}_{b \in \partial i \setminus a}) 
\ ,
\label{eq_n_atoi_gene}
\end{equation}
where the functions $\tf$ and $\tg$ are given by
\begin{eqnarray}
\left[\tf(\{ 
\vec{n}_{j \to a} \}_{j \in \partial a \setminus i}) \right]_{\t_i}
&\equiv&
\min_{\ut_{a \setminus i} } \left\{ 
\sum_{j \in \partial a \setminus i} 
[\vec{n}_{j \to a}]_{\t_j}
\ \Big| \  \psi_a(\t_i,\ut_{a \setminus i})=1 \right \} \ ,
\label{eq_def_tf}
\\
\left[\tg_\s(\vec{n}_1,\dots,\vec{n}_l)\right]_\t &\equiv& 
\I(\t \neq \s ) + [\vec{n}_1]_\t + \dots + [\vec{n}_l]_\t \ .
\label{eq_def_tg}
\end{eqnarray}
To lighten the notations we keep implicit the dependence of the functions
$\tf$ and $\tg$ on the edges of the factor graph.
These equations can be easily solved, for a given initial satisfying 
assignment $\us$, noting that the messages from the leaf variable nodes
$i$ satisfy the boundary condition $\vec{n}_{i \to a}(\s_i) = \vec{o}(\s_i)$,
where we define $[\vec{o}(\s)]_\t = \I(\s \neq \t)$. The recursions
(\ref{eq_n_atoi_gene}) can then be successively
applied to determine the value of all messages in a single sweep from the
exterior of the graph towards its center. When this is done the m.s.r. for
a variable $i$ is obtained from
\begin{equation}
\vec{n}_i(\us) =\tg_{\s_i}(\{ 
\vec{n}_{a \to i}(\s_i,\us_{a \to i}) \}_{a \in \partial i}) \ .
\label{eq_n_i_gene}
\end{equation}
Note that this recursive approach provides not only the size
of a minimal rearrangement, but also a final configuration achieving this 
bound. One just has to to bookkeep, along with the size informations
encoded in the messages $\vec{n}$, the configuration reaching the
minimum in Eq.~(\ref{eq_def_tf}) (if there are several of them one is
chosen arbitrarily). By construction the support of these 
optimal rearrangements is connected.

\subsubsection{Given tree, distribution with respect to $\us$}
\label{sec_gene_distri}
Following the program sketched above, we introduce now a probability 
distribution $\mu$ for the initial solution $\us$ of the formula:
\begin{equation}
\mu(\us) = \frac{1}{Z} \prod_{a} \psi_a(\us_a) 
\prod_{i \in B} \eta_{{\rm ext},i}(\s_i) \ ,
\label{eq_def_mu}
\end{equation}
where $Z$ is a normalization constant, $B$ is a subset of the leaves of the
factor graph, and the $\eta_{\rm ext}$ are probability laws on $\X$ that,
by analogy with magnetic systems, we shall call fields.
$\mu$ vanishes for configurations which do not satisfy the formula; if 
$B=\emptyset$ it is uniform on the set of solutions, otherwise the 
external fields $\eta_{\rm ext}$ can introduce a bias in the law (this
possibility will reveal useful in the following). We assume that the 
expression above remains well defined in the presence of the external fields,
i.e. that they do not put a vanishing weight on the solutions of the formula.

The absence of cycles in the factor graph induces a Markovian property
of the measure $\mu$ which greatly simplifies its characterization.
One can indeed compute recursively the marginals
of the law on any subset of variable nodes, introducing on each directed
edge of the factor graph another family of messages (cavity measures) 
$\nu_{a \to i}(\s_i)$ (resp. $\eta_{i \to a}(\s_i)$). 
These are the law of $\s_i$ in the measure associated to the cavity factor 
graph $F_{a \to i}$ (resp. $F_{i \to a}$), and are solutions of
\begin{eqnarray}
\nu_{a \to i} &=& f(\{\eta_{j \to a} \}_{j \in \partial a 
\setminus i}  ) \ \qquad f(\{\eta_{j \to a} \}_{j \in \partial a 
\setminus i}  ) (\sigma_i) =
\frac{1}{z(\{\eta_{j \to a} \}_{j \in \partial a 
\setminus i}  )}
\sum_{\us_{a \setminus i}} 
\psi_a(\s_i,\us_{a \setminus i}) \prod_{j \in \partial a \setminus i} 
\eta_{j \to a}(\sigma_j) \ , 
\label{eq_def_f} \\
\eta_{i \to a} &=& g(\{ \nu_{b \to i} \}_{b \in 
\partial i \setminus a}   ) \ \qquad g(\{ \nu_{b \to i} \}_{b \in 
\partial i \setminus a}   )  (\sigma_i) =
\frac{1}{z(\{ \nu_{b \to i} \}_{b \in 
\partial i \setminus a}   )}
\prod_{b \in \partial i \setminus a} \nu_{b \to i}(\sigma_i) \ ,
\label{eq_def_g}
\end{eqnarray}
where the functions $z$ are defined by normalization. Again for clarity
we do not indicate explicitly the dependence of the functions $f$, $g$ and 
$z$ on the edges.
The boundary conditions are $\eta_{i \to a}=\eta_{{\rm ext},i}$ 
when $i$ is a leaf in $B$, $\eta_{i \to a} = \oeta$ (the uniform
law on $\X$) if $i$ is a leaf not in $B$. 
This set of equations enjoys the same structure as the one on the 
$\vec{n}$'s (see Eq.~(\ref{eq_n_atoi_gene})), and can also be solved in a
sweep from the leaves of the factor graph. The marginals of $\mu$ for any
connected subset of variables can be easily expressed in terms of the
solution of this set of equations. For instance the marginal of a single
variable reads
\begin{equation}
\mu(\s_i) = g(\{ \nu_{a \to i} \}_{a \in \partial i})(\s_i ) \ ,
\end{equation}
while the variables of a constraint, conditioned to the value of one of them,
are drawn according to
\begin{equation}
\mu(\us_{a \setminus i} | \sigma_i ; 
\{ \eta_{j \to a} \}_{j \in \partial a \setminus i}  ) =
\frac{1}{z(\sigma_i,\{ \eta_{j \to a} \}_{j \in \partial a \setminus i})}
\psi_a(\s_i, \us_{a \setminus i}) \prod_{j \in \partial a \setminus i} 
\eta_{j \to a}(\sigma_j) \ ,
\label{eq_broadcast}
\end{equation}
where again $z$ is a normalizing factor.

We have now to compute the distribution of the minimal size rearrangements
when the starting configuration $\us$ is drawn from $\mu$. The generation of 
$\us$ can be performed in a recursive broadcasting way: one first
draws an arbitrarily chosen root variable $\s_i$ according to its
marginal $\mu(\s_i)$. Because the factor graph is a tree, the law of
the remaining variables factorizes on the different branches around $i$,
\begin{equation}
\mu(\us_{\setminus i} | \s_i) = \prod_{a \in \partial i} 
\mu(\us_{a \to i}| \s_i) \ .
\label{eq_breaking}
\end{equation}
For each branch $F_{a \to i}$ one proceeds by drawing the variables of
$\us_{a \setminus i}$, conditioned on $\s_i$ 
(see Eq.(\ref{eq_broadcast})). Then the value of $\s_j$ for each 
$j \in \partial a \setminus i$ conditions the generation of 
$\us_{j \to a}$, which can itself be broken in subtrees as in 
Eq.~(\ref{eq_breaking}). This process is repeated outwards until the leaves of
the tree are reached.

This observation leads us to introduce the distribution of the $\vec{n}$'s
messages with respect to the conditional distributions of the initial 
configuration,
\begin{equation}
q_{\vec{n}}^{(i \to a, \s_i )} = \sum_{\us_{i \to a} } 
\mu(\us_{i \to a}|\s_i)  
\ \delta_{\vec{n}, \vec{n}_{i \to a}(\s_i,\us_{i \to a})} \ , \qquad
\hq_{\vec{n}}^{(a \to i, \s_i )} = \sum_{\us_{a \to i} } 
\mu(\us_{a \to i}|\s_i)  
\ \delta_{\vec{n}, \vec{n}_{a \to i}(\us_{a \to i})} \ .
\end{equation}
Combining the recursive computations of the messages $\vec{n}$ expressed
in Eq.~(\ref{eq_n_atoi_gene}) and the recursive generation of the
initial configuration $\us$ leads to
\begin{eqnarray}
\hq^{(a \to i,\s_i)}_{\vec{n}} &=& \sum_{\us_{a \setminus i}} 
\mu(\us_{a \setminus i}  | \s_i ; \{ \eta_{j \to a} \} )
\prod_{j \in \partial a \setminus i} 
\sum_{\vec{n}_{j \to a}}
q^{(j \to a,\s_j)}_{\vec{n}_{j \to a}}
\ \delta_{\vec{n},\tf(\{\vec{n}_{j \to a}\}) } \ ,
\label{eq_broadcast2} \\
q^{(i \to a,\s_i)}_{\vec{n}} &=& 
\prod_{b \in \partial i \setminus a} 
\sum_{\vec{n}_{b \to i}}
\hq^{(b \to i,\s_i)}_{\vec{n}_{b \to i}}
\ \delta_{\vec{n},
\tg_{\s_i}(\{ \vec{n}_{b \to i} \} )} \ ,
\label{eq_q_itoa_gene}
\end{eqnarray}
with the boundary condition given by 
$q_{\vec{n}}^{(i \to a, \s_i)}=\delta_{\vec{n},\vec{o}(\s_i)}$
for the leaves $i$. The distribution of the m.s.r. for $i$ when $\us$ is 
drawn from $\mu$ can then be obtained from the distributions on the edges 
neighboring $i$,
\begin{equation}
q_n^{(i)} = \sum_{\s_i} \mu(\s_i) \sum_{\vec{n}} q_{\vec{n}}^{(i,\s_i)}
\ \delta_{n,\underset{\t_i \neq \s_i}{\min}[\vec{n}]_{\t_i}}
\ , \qquad
q_{\vec{n}}^{(i,\s_i)} = \prod_{a \in \partial i} 
\sum_{\vec{n}_{a \to i}}
\hq^{(a \to i,\s_i)}_{\vec{n}_{a \to i}}
\ \delta_{\vec{n}, \tg_{\s_i}(\{ \vec{n}_{a \to i} \} )} \ .
\label{eq_qn_i}
\end{equation}

\subsubsection{Average over the choice of the tree}
\label{sec_gene_ave}
At this point we define an ensemble of random rooted tree factor graphs on 
which we shall perform the average of the m.s.r. distribution. The ingredients 
of the definition are $p_l$, a distribution on the positive integers, 
$\rho(\psi)$ a distribution on the $0/1$ constraint functions (with possibly a 
random degree $k$), and a distribution of fields $\P(\eta)$. Let us denote 
$\T_L$ a random tree of the ensemble of depth $L$, and for notational 
simplicity $\hT_L$ the elements of this ensemble conditioned on their root 
being of degree one. $\T_L$ is defined by induction on $L$ as a (Galton-Watson 
like) branching process. $\T_0$ is made of a single 
variable node (the root) to which is applied an external field $\eta$ drawn 
from $\P$. $\hT_L$ is generated by introducing a root variable node $i$, 
connected to a single interaction node $a$ whose constraint function 
$\psi_a$ is drawn from $\rho$. Then each variable node in 
$\partial a \setminus i$ is taken to be the root of an independently 
generated $\T_L$. Conversely $\T_{L+1}$ is made by identifying the roots of 
$l$ (a random integer drawn from $p_l$) independent copies of $\hT_L$.

For each tree drawn from this ensemble the two recursive computations yield
a set of messages on each edge of the factor graph directed towards the root, 
$(\eta,\{ q_{\vec{n}}^{(\s)} \}_{\s =1}^q )$ for an edge from a variable to a 
constraint, 
$(\nu,\{ \hq_{\vec{n}}^{(\s)} \}_{\s =1}^q )$ from a constraint to a variable. 
The randomness in the
definition of the tree turn these objects into random variables, whose
distribution depends only on the distance between the considered edge and
the leaves. To be more precise, let us call
$\mathfrak{P}_L(\eta,\{ q_{\vec{n}}^{(\s)} \} )$ the distribution of 
$(\mu(\s_i),\{ q_{\vec{n}}^{(i,\s_i)} \}) $ when $i$ is the root of a random
$\T_L$ tree, and similarly 
$\widehat{\mathfrak{P}}_L(\nu, \{ \hq_{\vec{n}}^{(\s)} \})$
for the distribution of the messages directed to the root variable node of 
$\hT_L$.

One can first notice that the recursion between the messages $\eta,\nu$ do
not involve the size distributions $q_{\vec{n}}$ and $\hq_{\vec{n}}$, and
thus define $\P_L(\eta)$ as the marginal of $\mathfrak{P}_L$ disregarding 
the $q_{\vec{n}}$'s, and similarly $\hPP_L(\nu)$ from 
$\widehat{\mathfrak{P}}_L$. $\P_L$ and 
$\hPP_L$ obey functional equations of the form $\hPP_L = F[\P_L]$, 
$\P_{L+1} = G [\hPP_L ]$, with $\P_{L=0} = \P$, and where the functionals $F$ 
and $G$ have a compact distributional writing,
\begin{equation}
\nu \eqd f(\eta_1, \dots, \eta_{k-1},\psi) \ , \qquad
\eta \eqd g(\nu_1,\dots,\nu_l) \ .
\label{eq_RS}
\end{equation}
The first equation means that drawing a variable $\nu$ from $\hPP_L$ amounts
to drawing a constraint function $\psi$ from $\rho$, $k-1$ i.i.d. variables
$\eta_i$ from $\P_L$ and computing $\nu$ from Eq.~(\ref{eq_def_f}). Similarly
$\P_{L+1}$ is obtained from $\hPP_L$ thanks to Eq.~(\ref{eq_def_g}), with the
branching number $l$ drawn from $p_l$.
In the following we shall assume that the distribution $\P$ on the boundary
of the tree is a solution of the fixed point functional equation $\P=G[F[\P]]$.
This implies a stationarity property with respect to the number of
generation $L$, $\P_L=\P$, $\hPP_L=\hPP=F[\P]$. This justifies a posteriori the
choice we made of including non-trivial biases at the boundary in the law
(\ref{eq_def_mu}) : in generic models unbiased boundary conditions 
represented by $\P(\eta) = \delta(\eta - \oeta)$ do not satisfy this stationary
property, this will be in particular the case for the random $k$-SAT problem
studied below.

The evolution of the size distributions when iterating the tree
construction is coupled, through the term $\mu(\us_{a \setminus i}|\s_i)$
of Eq.~(\ref{eq_broadcast2}), to the $\eta,\nu$ messages. 
We are however interested in a
rather simple quantity, the average of the m.s.r. distribution of the
root (see Eq.~(\ref{eq_qn_i})) with respect to the random tree. It is thus
possible to compute an average of the $q_{\vec{n}}^{(i \to a,\s_i)}$ on
an edge of depth $L$, provided this average is \emph{conditioned}
on the value of the associated message $\eta_{i \to a}$. This conditional
average, denoted $q_{\vec{n}}^{(\s,L)}(\eta)$, and its counterpart
$\hq_{\vec{n}}^{(\s,L)}(\nu)$, are then found to obey the following equations,
\begin{eqnarray}
\hq_{\vec{n}}^{(\s,L)}(\nu) \hPP(\nu) &=&
\E_{\psi} \int d\P(\eta_1) \dots d\P(\eta_{k-1}) 
\ \delta(\nu - f(\eta_1,\dots,\eta_{k-1},\psi) )  \sum_{\s_1,\dots,\s_{k-1}}
\mu(\s_1,\dots,\s_{k-1}| \s,\eta_1,\dots,\eta_{k-1},\psi) \nonumber \\&&
\hspace{2cm}
\sum_{\vec{n}_1,\dots,\vec{n}_{k-1}} q_{\vec{n}_1}^{(\s_1,L)}(\eta_1) \dots
q_{\vec{n}_{k-1}}^{(\s_{k-1},L)}(\eta_{k-1}) 
\ \delta_{\vec{n},\tf(\vec{n}_1,\dots,\vec{n}_{k-1},\psi)} \ , 
\label{eq_hq_gene} \\
q_{\vec{n}}^{(\s,L+1)} (\eta) \P(\eta) &=& 
\sum_l p_l \int d\hPP(\nu_1) \dots d\hPP(\nu_l) 
\ \delta(\eta - g(\nu_1,\dots,\nu_l)) 
\sum_{\vec{n}_1,\dots,\vec{n}_l}
\hq_{\vec{n}_1}^{(\s,L)}(\nu_1) \dots \hq_{\vec{n}_l}^{(\s,L)}(\nu_l) 
\ \delta_{\vec{n},\tg_\s(\vec{n}_1,\dots,\vec{n}_l)} \ , \label{eq_q_gene}
\end{eqnarray}
with the boundary condition
$q_{\vec{n}}^{(\s,L=0)}(\eta)=\delta_{\vec{n},\vec{o}(\s)}$. Finally the 
sought-for average m.s.r.d. for the
root of a random tree of depth $L$ reads:
\begin{equation}
q_n^{(L)} = \int d\P(\eta) \sum_\s \eta(\s) \sum_{\vec{n}} 
q_{\vec{n}}^{(\s,L)}(\eta) \ 
\delta_{n,\underset{\t \neq \s}{\min} [\vec{n}]_\t} \ .
\label{eq_q_final_gene}
\end{equation}

The numerical resolution of Eqs.~(\ref{eq_hq_gene},\ref{eq_q_gene}) could
at first sight seem rather difficult, as they involve, for each value
of the random variable $\eta$ (or $\nu$), $q$ distributions of vectors 
$\vec{n}$. One can however devise a simple method, generalizing the
population dynamics algorithm of~\cite{MePa}. The important point is to
notice that for a given value of $\s$, 
$q_{\vec{n}}^{(\s,L)}(\eta)\P(\eta)$ can be viewed as a joint distribution
of variables $(\eta,\vec{n}^{(\s)})$, which can be numerically 
represented by a population of a large number $\cal N$ of couples 
$\{(\eta_i,\vec{n}^{(\s)}_i)\}_{i=1}^{\cal N}$. The empirical distribution
of these couples is taken as an approximation (known as a particle 
approximation in the statistics literature) of 
$q_{\vec{n}}^{(\s,L)}(\eta)\P(\eta)$. This suggests the following algorithm.
Initialize a population $\{ \eta_i \}_{i=1}^{\cal N}$ drawn i.i.d. from $\P$ 
(this shall be itself performed by a standard population dynamics approach), 
and associate to each of them $q$ vectors, $\vec{n}_i^{(\s)} = \vec{o}(\s)$.
We thus have, for trees of depth $L=0$, a population 
$\{(\eta_i,\vec{n}^{(1)}_i,\dots,\vec{n}^{(q)}_i)\}_{i=1}^{\cal N}$. To
take this population from depth $L$ to depth $L+1$ one has to
\begin{itemize}
\item[-]
generate in an i.i.d. way $\cal N$ elements 
$(\nu_j,\vec{n}^{(1)}_j,\dots,\vec{n}^{(q)}_j)$, with 
$j\in [{\cal N}+1,2{\cal N}]$ to avoid notational confusion, by:
\begin{itemize}
\item[$\bullet$] choosing randomly a constraint function $\psi$ from $\rho$, 
and $k-1$ indices $i_1,\dots,i_{k-1}$ uniformly at random in $[1,{\cal N}]$.

\item[$\bullet$] computing $\nu_j=f(\eta_{i_1},\dots,\eta_{i_{k-1}},\psi)$.

\item[$\bullet$] for each $\s \in [1,q]$, 

\begin{itemize}

\item generating a configuration $(\s_1,\dots,\s_{k-1})$ according to the law
$\mu(\cdot | \s, \eta_{i_1},\dots,\eta_{i_{k-1}},\psi )$.

\item computing $\vec{n}_j^{(\s)} = 
\tf(\vec{n}_{i_1}^{(\s_1)},\dots,\vec{n}_{i_{k-1}}^{(\s_{k-1})},\psi)$.

\end{itemize}

\end{itemize}

\item[-]
then generate a new population 
$\{(\eta_i,\vec{n}^{(1)}_i,\dots,\vec{n}^{(q)}_i)\}_{i=1}^{\cal N} $,
repeating for each $i \in [1,{\cal N}]$ independently the following steps :
\begin{itemize}
\item[$\bullet$] Choose randomly a degree $l$ from $p_l$
and $l$ indices $j_1,\dots,j_l$ uniformly at random in 
$[{\cal N}+1,2{\cal N}]$.

\item[$\bullet$] Compute $\eta_i=g(\nu_{j_1},\dots,\nu_{j_l})$.

\item[$\bullet$] For each $\s \in [1,q]$, compute $\vec{n}_i^{(\s)} = 
\tg_\s(\vec{n}_{j_1}^{(\s)},\dots,\vec{n}_{j_l}^{(\s)})$.

\end{itemize}
\end{itemize}

After $L$ iterations of these two steps, for a given value of $\s$,
an element $(\eta_i,\vec{n}_i^{(\s)})$ with $i$ uniformly chosen in 
$[1,{\cal N}]$ is distributed with the joint law 
$q_{\vec{n}}^{(\s,L)}(\eta)\P(\eta)$~\footnote{We do not claim that 
$(\eta_i,\vec{n}_i^{(1)},\dots,\vec{n}_i^{(q)})$ is drawn according
to $\P(\eta) q^{(1,L)}_{\vec{n}^{(1)}} \dots q^{(q,L)}_{\vec{n}^{(q)}}$,
i.e. that the $\vec{n}_i^{(\s)}$ are independent conditionally on $\eta_i$,
which is not true. The algorithm induces correlations between the various 
values of $\s$, yet these are irrelevant for the linear averages we compute.}.
We can thus complete the computation of $q_n^{(L)}$ in terms of a weighted 
histogram,
\begin{equation}
q_n^{(L)} = \frac{1}{\cal N} \sum_{i=1}^{\cal N} \sum_{\s=1}^q \eta_i(\s)
\ \delta_{n,\underset{\t \neq \s}{\min} [\vec{n}_i^{(\s)}]_\t} \ .
\end{equation}

We shall now examine how this general formalism can be applied to the three
exemplar problems of XORSAT, COL and SAT.

\subsection{$k$-XORSAT}

\subsubsection{On a given tree factor graph}

Let us recall the factor graph representation of a $k$-XORSAT formula we use:
the variables are Ising spins $\s_i=\pm 1$, and each constraint node $a$ is
satisfied if and only if the product of its $k$ neighboring variables 
$\prod_{i \in \partial a} \s_i$ is equal to a given constant 
$J_a = \pm 1$.
The computation of the m.s.r., already performed in~\cite{MoSe}, is much 
simpler than the general case presented above. Note first that for any CSP
where variable can only take two values,
a rearrangement $\us \to \ut$ is completely specified by its support, the set 
$R$ of variables which are different in the initial and final configurations. 
A second simplification is specific to the XORSAT problem. 
Consider an initial solution $\us$ and
the configuration $\ut$ obtained by flipping the variables in $R$. This
second configuration is also a solution if and only if for each constraint
$a$, an even (possibly) null number of variables of $\partial a$ are in $R$.
A rearrangement for the variable $i$ is hence a set $R$ verifying this 
condition and containing $i$. The m.s.r. $n_i$ is the minimal cardinality of
such a set of variables; on a tree this minimum can always be achieved
requiring that each $a$ contains either zero or two (and not an higher even 
value) variables of $R$. The recursive strategy for the computation of $n_i$ 
and the construction of a rearrangement of this size amounts to constructing a 
m.s.r. $R_{a \to i}$ for all the branches $F_{a \to i}$ around $i$ 
(their sizes being denoted $1 + n_{a \to i}$) and to combining the 
rearrangements of the sub-factor graphs, 
$R=\{ i \} \cup_{a \in \partial i} R_{a \to i}$. To construct $R_{a \to i}$
one has to choose exactly one variable $j \in \partial a \setminus i$
that minimizes the cost $n_{j \to a}$ of the rearrangement in the branch
$F_{j \to a}$. Summarizing this reasoning in formulas, we obtain:
\begin{equation}
n_{a \to i} = \min_{j \in \partial a \setminus i } n_{j \to a} \ , \qquad
n_{i \to a} = 1 + \sum_{b \in \partial i \setminus a} n_{b \to i}  \ , \qquad
n_i = 1+ \sum_{a \in \partial i} n_{a \to i} \ .
\label{eq_XORSAT}
\end{equation}
The reader will easily verify that the equations (\ref{eq_n_atoi_gene},\ref{eq_def_tf},\ref{eq_def_tg},\ref{eq_n_i_gene}) of
the general formalism reduce indeed to this simple form,
noting in particular that the m.s.r. is here independent of the initial 
configuration, as appears clearly from the geometric characterization
of the optimal supports $R$.

\subsubsection{Random tree}
\label{sec_xsat_random}

This independence with respect to the initial configuration allows to skip
the second step of the general formalism, as for a given tree the distribution
of the m.s.r. is trivially concentrated on a single integer, and to
study directly the ensemble of random tree formula. We shall follow
the general definition of $\T_L$ given above, with a Poisson law of parameter
$\alpha k$ for the branching probability $p_l$, and all constraint nodes
of degree $k$. For definiteness one can assume that the boundary condition
is free (no bias on the leaves of the tree) and that $J_a = \pm 1$ with
probability one half; these last two choices are in fact irrelevant, as the 
m.s.r. depends only on the geometry of the factor graph.

This random ensemble induces a probability law $q_n^{(L)}$ for the m.s.r. 
of the root of $\T_L$, and an associated law $\hq_n^{(L)}$ for the message sent
to the root of $\hT_L$. Simplifying the
equations~(\ref{eq_hq_gene},\ref{eq_q_gene},\ref{eq_q_final_gene}) of the 
general formalism, or interpreting the specific ones
(\ref{eq_XORSAT}) in a distributional sense, leads to 
\begin{eqnarray}
\hq_n^{(L)} &=& \sum_{n_1,\dots,n_{k-1}} q_{n_1}^{(L)} \dots q_{n_{k-1}}^{(L)} 
\ \delta_{n,\min[n_1,\dots,n_{k-1}]} \ ,
\label{eq_hq_XORSAT} \\
q_n^{(L+1)} &=& \sum_{l=0}^\infty \frac{e^{-\alpha k} (\alpha k)^l}{l!}
\sum_{n_1,\dots,n_l} \hq_{n_1}^{(L)} \dots \hq_{n_l}^{(L)} 
\ \delta_{n,1+n_1+\dots+n_l} \ ,
\label{eq_q_XORSAT} 
\end{eqnarray}
with the initial condition $q_n^{(L=0)} = \delta_{n,1}$.

These equations can be solved by a simplified version of the population 
dynamics algorithm introduced in the general case. The distributions 
$q_n^{(L)}$ and $\hq_n^{(L)}$ are represented by samples of integers 
$\{n_i\}$, each element of the population associated to 
$q_n^{(L+1)}$ is generated by drawing a
Poisson distributed integer $l$, extracting at random $l$ elements of the 
sample representing $\hq_n^{(L)}$ and computing their sum plus one. Conversely
the elements of $\hq_n^{(L)}$ are the minimum of $k-1$ randomly chosen integers
drawn from the population encoding $q_n^{(L)}$. 
In the following we shall be interested in
the $L \to \infty$ limit, which is the counterpart of the $N \to \infty$ 
thermodynamic limit of the original random graph ensembles. One could reach
it numerically by repeated iterations of the population dynamics step.
There is however a simpler numerical method which allows to 
perform analytically this limit.

Let us first define
the integrated version of the m.s.r.d.,
\begin{equation}
Q_n^{(L)} = \sum_{n' \ge n} q_{n'}^{(L)} \ ,
\end{equation}
which gives the probability of a m.s.r. being larger than $n$. A few
simple properties follow from this definition,
\begin{equation}
q_n^{(L)} = Q_n^{(L)} - Q_{n+1}^{(L)} , \qquad
Q_n^{(L)} = 1 - \sum_{n'<n} q_{n'}^{(L)} , \qquad
\lim_{n \to \infty} Q_n^{(L)} = 0 \ . 
\label{eq_prop_Qn}
\end{equation}
A slightly less obvious property is that, for a fixed value of $n$, 
$Q_n^{(L)}$ is monotonously increasing with $L$.
This arises from the fact that larger trees have larger rearrangements,
and can be proven from (\ref{eq_hq_XORSAT},\ref{eq_q_XORSAT}) via a standard 
stochastic domination argument~\cite{coupling}.
Being moreover bounded from above by $1$, $Q_n^{(L)}$ converges as $L$ goes to
infinity, to a limit we shall denote $Q_n$. By continuity in the first equality
of (\ref{eq_prop_Qn}) the limit $q_n$ of $q_n^{(L)}$ also exists; same 
statements apply to $\hQ_n$ and $\hq_n$. Eq.~(\ref{eq_hq_XORSAT}) can
be rewritten as $\hQ_n^{(L)} =(Q_n^{(L)})^{k-1}$, in the infinite $L$ limit
we thus obtain:
\begin{eqnarray}
\hQ_n &=& Q_n^{k-1} \ ,
\label{eq_hq_XORSAT_Linfty} \\
q_n &=& \sum_{l=0}^\infty \frac{e^{-\alpha k} (\alpha k)^l}{l!}
\sum_{n_1,\dots,n_l} \hq_{n_1} \dots \hq_{n_l} 
\ \delta_{n,1+n_1+\dots+n_l} \ . 
\label{eq_q_XORSAT_Linfty} 
\end{eqnarray}
These limit distributions can now be determined by a recursion on $n$.
Eq.~(\ref{eq_q_XORSAT}) implies that $q_1^{(L)}=e^{-\alpha k}$ for all 
$L \ge 1$; hence $q_1=e^{-\alpha k}$,
which fixes the starting point of the recursion. Assume
$q_n$ has been computed upto rank $m$. This means that 
$Q_n = 1 - \sum_{n'<n} q_{n'}$
is known upto rank $m+1$, and the same is true for $\hQ_n$ because
of Eq.~(\ref{eq_hq_XORSAT_Linfty}).
We thus have at our disposal the values of $\hq_n$ upto $n=m$, which
allows the computation of $q_{m+1}$ through Eq.~(\ref{eq_q_XORSAT_Linfty}).
We defer the presentation of the numerical results obtained in this way
until Section~\ref{sec_results}, in order to confront them with the COL and
SAT problems.

Let us only anticipate one feature by emphasizing that the limit 
$L \to \infty$ was taken here at a fixed value of $n$. We shall see
that for some values of $\alpha$ the limits $L,n \to \infty$ do not
commute, a situation reminiscent of a percolating regime.
In such cases $Q_n$ tends for large $n$ to a strictly positive
value $\phi$, $q_n$ is not 
normalized anymore and cannot be directly considered as the distribution
of an integer random variable $n$. It will be however convenient to formally 
consider $n$ as an extended integer, with a probability $\phi$ of being 
infinite.

\subsection{$q$-COL}

\subsubsection{Given tree, given $\us$}

The second example of CSP we shall consider is the $q$-coloring problem.
The variables $\s_i$ can take one of the $q$ values (colors) in
$\{1,\dots,q\}$, and the constraint node $a$ linking two variables
$i$ and $j$ forbids the configurations with $\s_i = \s_j$. The
solutions of this CSP are thus the proper colorings of the underlying graph.

At variance with the XORSAT problem, the m.s.r. 
does depend on the initial satisfying assignment: take for
instance a small graph made of a central site $i$ with $q-1$ neighbors. If
in the initial coloring all the peripheral sites have distinct colors, 
the minimal size to rearrange $i$ is two. Otherwise, if at least two 
peripheral sites have the same color, there is one color available for
the central site to be rearranged without modifying its neighborhood.

There is however room for simplifications with respect to the general 
formalism. Consider the constraint $a$ between two adjacent vertices $i$ and
$j$. The vectorial message $\vec{n}_{a \to i}(\us_{a \to i})$ has only
one non-zero component, corresponding to the perturbation 
$\s_i \to \t_i=\s_j$. 
This is a formal consequence of Eq.~(\ref{eq_def_tf}), 
but has a very intuitive meaning: in the cavity graph $F_{a \to i}$ the root 
$\s_i$ can be given any value $\t_i \neq \s_j$ without having to propagate the 
rearrangement.
We can thus get rid of the vectorial character of the messages. Note also
that the information contained in the messages $\vec{n}_{a \to i}$ and 
$\vec{n}_{j \to a}$ is redundant, as each constraint node involves
only two variables. We shall thus eliminate the variable to constraint 
messages, and rename $n_{j \to i}(\us_{j \to i})$ what was denoted
in the general formalism $[\vec{n}_{a \to i}(\us_{a \to i})]_{\s_j}$.
Simplifying Eqs.~(\ref{eq_n_atoi_gene},\ref{eq_def_tf},\ref{eq_def_tg},\ref{eq_n_i_gene}) with these new notations, we obtain
\begin{eqnarray}
n_{j\to i}(\us_{j \to i}) &=& 1 + \min_{\t_j \neq \s_j} \left\{ 
\sum_{k\in \partial j \setminus i}  
\delta_{\s_k,\t_j} n_{k \to j}(\us_{k \to j})   \right\} \ ,
\label{eq_n_itoj_col} \\
n_i(\us) &=& 1 + \min_{\t_i \neq \s_i} \left\{ 
\sum_{j \in \partial i}  
\delta_{\s_j , \t_i} n_{j \to i}(\us_{j \to i})   \right\} \ ,
\label{eq_n_i_col}
\end{eqnarray}
with $n_{j \to i}(\s_j) =1$ if $j$ is a leaf of the tree.
The interpretation of these equations is clear: to modify the color
$\s_i$ of a vertex $i$ in a coloring $\us$ one has to probe the $q-1$ 
possibilities of $\t_i \neq \s_i$, and follow the effect of this modification
in the branches $F_{j \to i}$ that become unsatisfied, i.e. those
who had $\s_j = \t_i$ before the modification.

\subsubsection{Given tree, distribution with respect to $\us$}

We shall study in the coloring case the distribution of the m.s.r. with
respect to the measure $\mu(\us)$ uniform on the proper colorings.
In other words we use a free boundary condition and do not impose any 
external field on the leaves. This choice preserves the permutation 
symmetry among colors, which implies that the marginal distribution 
$\mu(\s_i)$ of any variable $i$ is uniform over the $q$ possible values.
Once the color of an arbitrary root variable $i$ has been chosen,
the generation of the remaining sites can be done in a recursive way: the
colors of the neighbors of $i$ are drawn independently, uniformly over the
$q-1$ colors distinct from $\s_i$, and this process is repeated from $i$ 
outwards.
Exploiting this symmetry and the recursions 
(\ref{eq_n_itoj_col},\ref{eq_n_i_col}), one finds that the distributions of
the m.s.r. with respect to the uniform choice of the initial proper coloring
is given by
\begin{equation}
q_{n}^{(i)} = \frac{1}{(q-1)^{|\partial i|}} \prod_{j \in \partial i}
\sum_{\substack{ \s_j \neq 1 \\ n_{j \to i}}  } q_{n_{j \to i}}^{( j \to i)}
\ \ \I\left(n=1 + \min_{\s \neq 1 } \left[\sum_{j \in \partial i} 
\delta_{\s,\s_j} n_{j \to i} \right] \right) \ ,
\label{eq_qn_col}
\end{equation}
where the distributions of the messages on the edges of the tree are solutions
of
\begin{equation}
q_n^{(j \to i)} = \frac{1}{(q-1)^{|\partial j|-1}} 
\prod_{k \in \partial j \setminus i}
\sum_{\substack{ \s_k \neq 1 \\ n_{k \to j}}  } 
 q_{n_{k \to j}}^{( k \to j)}
\ \ \I\left(n=1 + \min_{\s \neq 1 } \left[\sum_{k \in \partial j \setminus i} 
\delta_{\s,\s_k} n_{k \to j} \right] \right) \ ,
\label{eq_qn_jtoi_col}
\end{equation}
with the boundary condition $q_n^{(j \to i)}=\delta_{n,1}$ when $j$ is a leaf.

\subsubsection{Average over the choice of the tree}

We now consider the ensemble of random trees $\T_L$ where the variable nodes
have a Poissonian branching probability of mean $c$, and all constraint
nodes are identical, $\psi(\s_i,\s_j)=\I(\s_i \neq \s_j)$.
One can easily show from Eqs.~(\ref{eq_qn_col},\ref{eq_qn_jtoi_col}) that
the m.s.r.d. for uniformly distributed initial proper colorings,
averaged over this random tree ensemble is given by
\begin{equation}
q_n^{(L+1)} = \sum_{l=0}^\infty \frac{e^{-c}c^l}{l!} \frac{1}{(q-1)^l}
\sum_{\s_1,\dots,\s_l = 2}^q
\sum_{n_1,\dots,n_l} q_{n_1}^{(L)} \dots q_{n_l}^{(L)}  \ \ 
\I\left(n= 1 + \min_{\s=2,\dots,q} 
\left[\sum_{i=1}^l \delta_{\s,\s_i} n_i \right] \right) \ ,
\label{eq_qn_av_col}
\end{equation}
with $q_n^{(L=0)}=\delta_{n,1}$. This equation could be solved following
the population dynamics approach explained above. One can however unveil
a formal equivalence with the computation performed for the XORSAT
problem. Consider indeed the random variables $l_\s$ which counts
in Eq.~(\ref{eq_qn_av_col}) the number of $\s_i$'s assigned to the value $\s$.
Conditional on $l$ the $l_\s$'s are multinomially distributed; as $l$ is
itself a Poisson random variable the $l_\s$ turn out to be independent Poisson
random variables. This allows to rewrite Eq.~(\ref{eq_qn_av_col}) as
\begin{equation}
q_n^{(L+1)} =\sum_{l_2,\dots,l_q = 0}^\infty \frac{e^{-c} 
\left(\frac{c}{q-1}\right)^{l_2 + \dots + l_q} }{l_2! \dots l_q!} 
\sum_{m_2,\dots,m_q} \delta_{n,\min[m_2,\dots,m_q]} 
\prod_{\s=2}^q \left( \sum_{n_\s^1,\dots,n_\s^{l_\s} } 
q^{(L)}_{n_\s^1} \dots 
q^{(L)}_{n_\s^{l_\s}} \delta_{m_\s,1+n_\s^1+\dots+n_\s^{l_\s}} \right)
\ .
\end{equation}
Comparing with Eqs.~(\ref{eq_hq_XORSAT},\ref{eq_q_XORSAT})
one realizes that the solution of the coloring case can
be directly read off from the study of the XORSAT one with a simple
translation of the parameters,
\begin{equation}
q_n^{(L,{\rm COL})}[q,c] = \hq_n^{(L,{\rm XORSAT})}
\left[k=q,\alpha=\frac{c}{q(q-1)} \right] \ .
\label{eq_traduction}
\end{equation}
In particular the simple recursion on $n$ to solve directly in the 
$L \to \infty$ limit is still applicable to the coloring problem.

\subsection{$k$-SAT}

\subsubsection{Given tree, given $\us$}

We consider now the third example of CSP, in which the factor graph encodes
a $k$-satisfiability formula. The boolean variables are represented by Ising 
spins $\s_i = \pm 1$; each constraint node $a$ is linked to $k$ variable
nodes, and is unsatisfied if and only if these $k$ variable all takes their 
unsatisfying value, $\s_i = J_i^a$ for all $i \in \partial a$.
We shall denote $\partial_+ i(a)$ (resp. $\partial_- i(a)$) the set of clauses
in $\partial i \setminus a$ agreeing (resp. disagreeing) with $a$ on the
satisfying value of $\s_i$. 
We also denote $\partial_\sigma i$ the set of clauses
in $\partial i$ which are satisfied by $\sigma_i=\sigma$.

Because of the boolean nature of the variables a rearrangement is specified
by the set of variables to be flipped (recall the discussion of the XORSAT
problem), we can get rid of the vectorial character of the general
formalism and denote, for instance, $n_i(\us)$ for the m.s.r. of the variable
$i$ under the perturbation $\s_i \to \t_i = - \s_i$. This quantity does 
depend on the initial satisfying assignment. In the simplest case where there 
is one single constraint node $a$ in the factor graph, $n_i(\us)=2$ if
$a$ was satisfied only by $i$ before its flip, $n_i(\us)=1$ for
all the other satisfying assignments.
Generalizing this observation to generic factor graphs, one reduces
the recursion relations of the general formalism (see 
Eqs.~(\ref{eq_n_atoi_gene},\ref{eq_def_tf},\ref{eq_def_tg},\ref{eq_n_i_gene})) 
to:
\begin{eqnarray}
n_{a \to i}(\s_i,\us_{a \to i}) &=& \begin{cases} 
\underset{j \in \partial a \setminus i}{\min} n_{j \to a}(\s_j,\us_{j \to a}) 
& \mbox{if $\s_i = -J_i^a$ and $\s_j = J_j^a \ \ \ \forall j \in \partial a 
\setminus i$}\\
0 & \mbox{otherwise}
\end{cases} \ , \\
n_{i \to a}(\s_i,\us_{i \to a}) &=& 1 + \sum_{b \in \partial i \setminus a} 
n_{b \to i}(\s_i,\us_{b \to i}) \ , \\
n_i(\us) &=& 
1 + \sum_{a \in \partial i} n_{a \to i}(\s_i,\us_{a \to i}) \ ,
\end{eqnarray}
with again $n_{i \to a}(\s_i)=1$ for the leaves of the graph.

\subsubsection{Given tree, distribution with respect to $\us$}

We now consider the probability law $\mu(\us)$ on the initial satisfying
assignments, with external fields on some of the leaves of the graph.
More precisely, we use the form (\ref{eq_def_mu}), with the biases on a subset
$B$ of the leaves parameterized by a real $h_{{\rm ext},i}$:
\begin{equation}
\eta_{{\rm ext},i}(\s_i) = \frac{1+\s_i \tanh h_{{\rm ext},i}}{2} \ .
\end{equation}
The messages $\nu_{a \to i}$ and $\eta_{i \to a}$ are probability laws of
Ising spins and can thus be parameterized by a single real. To simplify
the notations we make a gauge transformation with respect
to the value of the variable satisfying the clause and define
\begin{equation}
\nu_{a \to i}(\s_i) = \frac{1- J_a^i \s_i \tanh u_{a \to i}}{2} \ , \qquad
\eta_{i \to a}(\s_i) = \frac{1 - J_a^i \s_i \tanh h_{i \to a}}{2} \ .
\end{equation}
With these conventions Eqs.~(\ref{eq_def_f},\ref{eq_def_g}) become
\begin{eqnarray}
u_{a \to i} &=& f(\{ h_{j \to a} \}_{j\in \partial a \setminus i} ) \ , \qquad
f(h_1,\dots,h_{k-1}) =
-\frac{1}{2} \ln \left( 1-  \prod_{i=1}^{k-1} \frac{1-\tanh h_i}{2} \right)
\ , \\
h_{i \to a} &=& \sum_{b \in \partial_+ i(a) } u_{b \to i}
- \sum_{b \in \partial_- i(a) } u_{b \to i} \ ,
\end{eqnarray}
with $h_{i \to a} = - J_i^a h_{{\rm ext},i}$ if $i$ is a leaf in $B$, 0 
if it is a leaf not in $B$. 
The solution of these equations allows to compute the
two quantities that we shall need below:
\begin{itemize}
\item[$\bullet$] the marginal law of $\s_i$,
\begin{equation}
\mu(\s_i) = \frac{1 + \s_i \tanh h_i}{2} \ , \qquad
h_i =\sum_{a \in \partial_+ i } u_{a \to i} - 
\sum_{a \in \partial_- i } u_{a \to i} \ .
\end{equation}

\item[$\bullet$] the probability that, conditional on $\s_i$ satisfying
the constraint $a$, all other variables in $\partial a$ take their
wrong values,
\begin{equation}
\mu( \s_j = J_j^a \ \ \forall j \in \partial a \setminus i | \s_i=-J_i^a) = 
\prod_{j \in \partial a \setminus i} \frac{1 - \tanh h_{j \to a}}{2}
\end{equation}

\end{itemize}

We now proceed with the introduction of the distributions 
$\hq_n^{(a \to i,\s_i)}$ (resp. $q_n^{(i \to a,\s_i)}$) of the messages
$n_{a \to i}(\s_i, \us_{a \to i})$ (resp. $n_{i \to a}(\s_i, \us_{i \to a})$)
when $\us_{a \to i}$ (resp. $\us_{i \to a})$) is drawn conditionally on
$\s_i$. In fact for each directed edge the 
distribution corresponding to one of the two values of $\s_i$ can be discarded.
Consider first the cavity factor graph $F_{a \to i}$. If $\us_{a \to i}$
is drawn conditionally on $\s_i$ not satisfying constraint $a$, necessarily
one of the $k-1$ other variables of $a$ will satisfy it so that 
$\s_i$ can be flipped without propagating the rearrangement further in the 
branch. This is translated in formula as 
$\hq_n^{(a \to i,\s_i = J_i^a)} = \delta_{n,0}$, we shall
thus simplify notation and write $\hq_n^{(a \to i )}$ instead of
$\hq_n^{(a \to i,\s_i = -J_i^a )}$ for the only non-trivial size distribution
born by the edge $a \to i$. This last quantity, in virtue of
Eq.~(\ref{eq_broadcast2}), has to be expressed in terms of the distributions
$q_n^{(j \to a, \s_j)}$ for $j \in \partial a \setminus i$. However the
rearrangement has to be propagated only if none of these variables were
satisfying constraint $a$, we can thus rename 
$q_n^{(j \to a)} \equiv q_n^{(j \to a, \s_j = J_j^a)}$ and forget about
$q_n^{(j \to a, \s_j = - J_j^a)}$. Collecting these various observations
we obtain
\begin{eqnarray}
\hq_n^{(a \to i )} &=& \prod_{j \in \partial a \setminus i}
\sum_{ n_{j \to a} } q_{n_{j \to a}}^{(j \to a)}
\left[
\left(1 - \prod_{j \in \partial a \setminus i} 
\frac{1 - \tanh h_{j \to a}}{2} \right) \delta_{n,0} +
\left(\prod_{j \in \partial a \setminus i} \frac{1 - \tanh h_{j \to a}}{2}
\right)
\ \delta_{n,\min \{ n_{j \to a} \} }
\right] \ , \\
q_n^{(i \to a)} &=& 
\prod_{b \in \partial_-i(a)}
\sum_{ n_{b \to i} } \hq_{n_{b \to i} }^{(b \to i)}
\ \delta_{n,1+\sum n_{b \to i}} \ ,
\end{eqnarray}
with $q_n^{(i \to a)} = \delta_{n,1}$ on the leaves. Finally the law
of the m.s.r. for $i$ is given by
\begin{equation}
q_n^{(i)} = \sum_\s \frac{1+ \s \tanh h_i}{2}
\prod_{a \in \partial_\sigma i}
\sum_{n_{a \to i}}
 \hq_{n_{a \to i}}^{(a \to i)} 
\ \delta_{n,1+\sum n_{a \to i}} \ .
\label{eq_qni_sat}
\end{equation}

\subsubsection{Average over the choice of the tree}

We shall study random trees $\T_L$ with a Poissonian law of mean $\alpha k$ 
for the branching probability $p_l$ of variable nodes. The constraint nodes
are all of degree $k$ with the signs $J_i^a$ of the unsatisfying literals 
i.i.d. random variables equal to $\pm 1$ with equal probability. This implies
that the cardinality of the neighborhoods $\partial_+ i$ and $\partial_- i$ of
the root are two independent Poisson random variables of mean $\alpha k/2$,
whose law shall be denoted $p_{l_+,l_-}$. The same statement is true for 
$\partial_+i(a)$ and $\partial_- i(a)$ in the bulk of the tree. The last
element defining $\T_L$ is the distribution $\P(h)$ for the biases on the
leaves of depth $L$  of the tree. Following the general formalism we
assume this distribution to be stationary under the iterations
\begin{equation}
u \eqd f(h_1,\dots,h_{k-1}) \ , \qquad 
h \eqd \sum_{i=1}^{l_+} u_i^+ - \sum_{i=1}^{l_-} u_i^- \ ,
\label{eq_sat_RS}
\end{equation}
where $l_\pm$ are drawn from $p_{l_+,l_-}$ and the $h_i$ (resp. the $u_i^\pm$) 
are independent copies drawn from $\P(h)$ (resp. $\hPP(u)$).
The computation proceeds with the introduction of $q_n^{(L)}(h)$ 
(resp. $\hq_n^{(L)}(u)$),
the average of the $q_n^{(i \to a)}$ (resp. $\hq_n^{(a \to i)}$) conditioned
by the event $h_{i \to a} = h$ (resp. $u_{a \to i}=u$). The generic equations
(\ref{eq_hq_gene},\ref{eq_q_gene}) translate into
\begin{eqnarray}
\hq^{(L)}_n(u) \hPP(u) &=& \int \prod_{i=1}^{k-1} d\P(h_i) 
\ \delta(u-f(h_1,\dots,h_{k-1})) 
\label{eq_hq_sat} \\ && \hspace{8mm} 
\sum_{n_1,\dots,n_{k-1}} \prod_{i=1}^{k-1} q^{(L)}_{n_i}(h_i)\left[ 
 \left(1 - \prod_{i=1}^{k-1} \frac{1-\tanh h_i}{2} \right) \delta_{n,0}
+ \left (\prod_{i=1}^{k-1} \frac{1-\tanh h_i}{2} \right) 
\ \delta_{n,\min [n_1,\dots,n_{k-1}]}
\right] \ , \nonumber \\
q_n^{(L+1)}(h) \P(h) &=& \sum_{l_+,l_-=0}^\infty 
p_{l_+,l_-}
\int \prod_{i=1}^{l_+} d\hPP(u_i^+)
\prod_{i=1}^{l_-} d\hPP(u_i^-) \ 
\delta\left(h -\sum_{i=1}^{l_+} u_i^+ + \sum_{i=1}^{l_-} u_i^- \right)
\label{eq_q_sat}\\ && \hspace{8mm}
\sum_{n_1,\dots,n_{l_-}} \prod_{i=1}^{l_-} \hq^{(L)}_{n_i}(u_i^-) \ 
\delta_{n,1+n_1+\dots+n_{l_-} } \ , \nonumber 
\end{eqnarray}
with $q_n^{(L=0)}(h)=\delta_{n,1}$.
Finally the sought-for average m.s.r.d. for the root of $\T_L$ reads
\begin{equation}
q_n^{(L)} = \int d\P(h) \ (1 - \tanh h) \ q^{(L)}_n(h) \ ,
\label{eq_sat_final}
\end{equation}
which is obtained from (\ref{eq_qni_sat}) by using the statistical 
equivalence between positive and negative literals. This implies in 
particular that $h$ has a symmetric distribution, so that $q_n^{(L)}$ is well
normalized.

The adaptation of the general population dynamics algorithm to this case
is simple. The joint distribution $q_n^{(L)}(h)\P(h)$ 
is represented by a sample
of couples $\{(h_i,n_i)\}_{i=1}^{\cal N}$, initialized with $n_i=1$ and
the $h_i$'s distributed according to $\P(h)$ 
(thanks to preliminary population dynamics steps). The recursion over
$L$ amounts to generating a sample $\{(u_j,n_j)\}$, where for each $j$ 
one selects $k-1$ indices $i_1,\dots,i_{k-1}$ in $[1,{\cal N}]$. 
$u_j$ is set to $f(h_{i_1},\dots,h_{i_{k-1}})$, $n_j$ to the minimum
of $\{n_{i_1},\dots,n_{i_{k-1}} \}$ with probability $1-\exp[-2 u_j]$, to 0
otherwise. In the converse step for each new element $i$ two Poisson integers
$l_\pm$ of mean $\alpha k/2$ are independently drawn, then two sets of indices 
$J_+$ and $J_-$ of cardinalities $l_+$ and $l_-$ are generated. 
$h_i$ is given by $\sum_{j \in J_+} u_j - \sum_{j \in J_-} u_j$, while $n_i$
reads $1 + \sum_{j \in J_-} n_j$. From (\ref{eq_sat_final}) we obtain
$q_n^{(L)}$ as a weighted histogram of the population,
\begin{equation}
q_n^{(L)} = 
\frac{1}{\cal N} \sum_{i=1}^{\cal N} (1-\tanh h_i) \ \delta_{n,n_i} \ . 
\end{equation}
The large $L$ limit is obtained by repeating a sufficient number of these 
steps to achieve convergence within numerical precision.

\section{The freezing transition in random tree ensembles}
\label{sec_results}

In the previous section we have established numerical procedures to
compute the average m.s.r.d. $q_n$ for the various random tree ensembles,
based either on a simple recurrence over $n$ for the XORSAT and COL case,
or on a more elaborate population dynamics algorithm for SAT. We now discuss
the outcome of these computations, the limit of infinite depth trees
($L \to \infty$) being kept implicit. 
\begin{figure}
\begin{center}
\includegraphics[width=8cm]{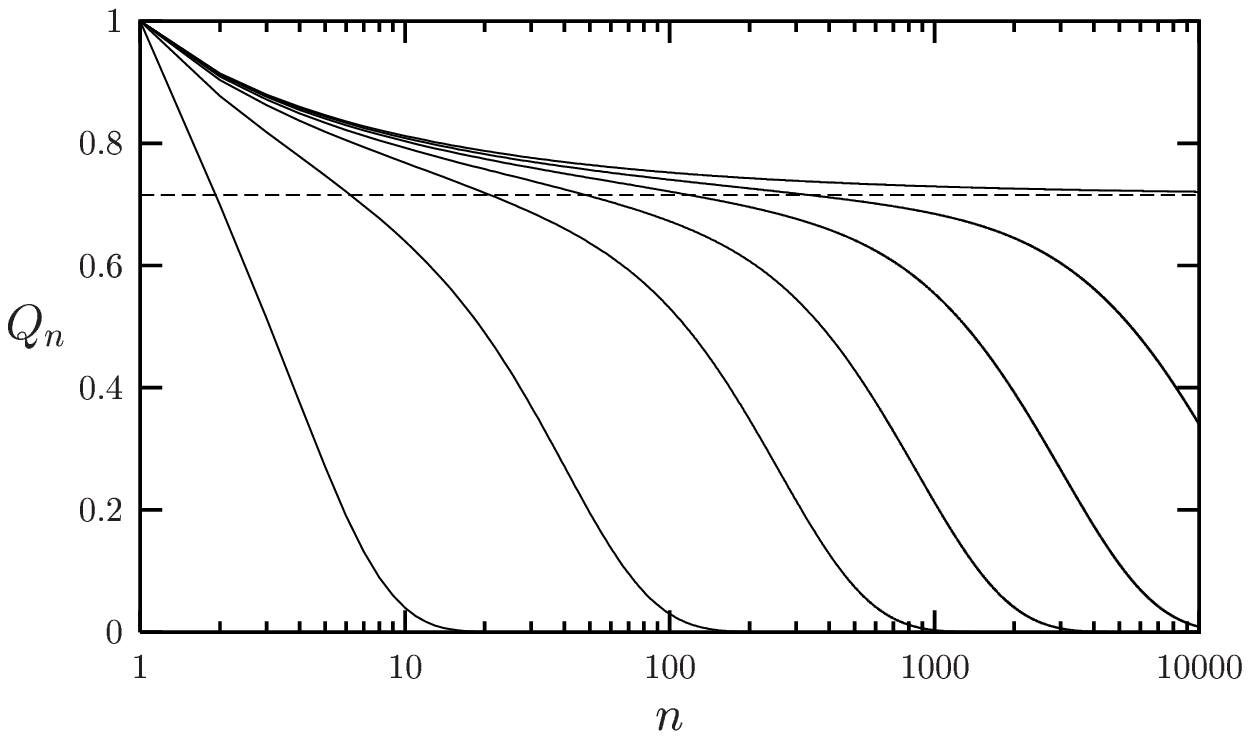} \hspace{1cm}
\includegraphics[width=8cm]{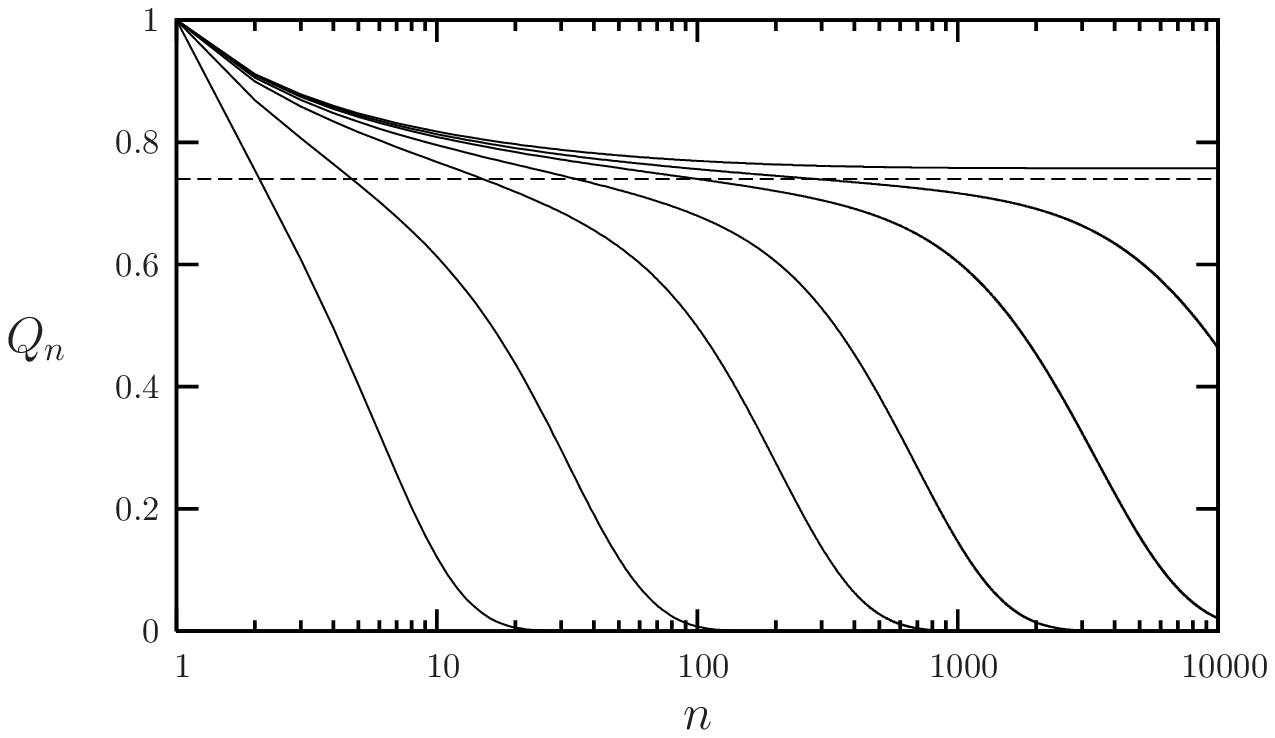}
\end{center}
\caption{Integrated average distribution of minimal size rearrangements
in tree ensembles. 
Left: random 3-XORSAT, from left to right 
$\alpha= 0.4$, $0.7$, $0.78$, $0.8$, $0.81$, $0.815$, $\alpha_{\rm p}$.
The dashed horizontal line is the order parameter at the transition, 
$\phi_{\rm p} \approx 0.715332$. 
Right: random 3-SAT, from left to right,
$\alpha=3$, $4$, $4.3$, $4.36$, $4.39$, $4.40$, $4.41$. The dashed line
indicates $\phi_{\rm p} \approx 0.74$.}
\label{fig_results}
\end{figure}

In Fig.~\ref{fig_results} we have plotted the integrated distribution $Q_n$,
for various values of $\alpha$, in the XORSAT and SAT case. 
These two families of curves present the same striking feature: when $\alpha$
is increased $Q_n$ develops a plateau, in other words $q_n$ becomes bimodal
with a positive fraction of rearrangements shifting towards larger and larger
values. When a critical value $\alpha_{\rm p}$ is reached the length of
the plateau becomes infinite. This transition is thus described by the
order parameter $\phi=\lim_{n \to \infty} Q_n$, which represents
the fraction of percolating optimal rearrangements whose size diverge with
$L$. From 
the point of view of the order parameter the transition is discontinuous,
$\phi$ jumps from $0$ to a positive value $\phi_{\rm p}$ when the threshold
$\alpha_{\rm p}$ is crossed. 

Let us follow the interpretation suggested at the end of 
Sec.~\ref{sec_xsat_random} of $Q_n$ being the distribution of an extended
integer which has probability $\phi$ of being infinite. With the rules
that the minimum of several such extended integers is infinite if and only if
each of them is infinite, while their sum is infinite as soon as one of them
is so, Eqs.~(\ref{eq_hq_XORSAT_Linfty},\ref{eq_q_XORSAT_Linfty}) 
imply in the XORSAT case
\begin{equation}
\hphi = \phi^{k-1} \ , \qquad \phi= 1 - \exp[-\alpha k \hphi ] \ ,  
\label{eq_param_xor}
\end{equation}
where we denoted $\hphi = \lim \hQ_n$. This can be closed under the form
$\phi= 1 - \exp[-\alpha k \phi^{k-1}]$, with $\alpha_{\rm p}$ being the 
smallest value of $\alpha$ for which there exists a non-trivial solution. 
At $\alpha_{\rm p}$ this solution appears discontinuously, with the positive
value $\phi_{\rm p}$ corresponding to the height of the plateau in the curves
of $Q_n$. For larger values of $\alpha$ there are two non-trivial solutions, 
the relevant one being the largest. Numerical values of $\alpha_{\rm p}$ and 
$\phi_{\rm p}$ are given in Table~\ref{tab} for a few values of $k$.

Thanks to the formal equivalence between XORSAT and COL summarized in
Eq.~(\ref{eq_traduction}) we immediately obtain the equation on the order
parameter of the COL freezing transition and the critical value $c_p$ (see
also Table~\ref{tab} for their numerical values), 
\begin{equation}
\phi = \left(1- \exp\left[ -\frac{c \phi}{ (q-1)} \right]\right)^{q-1} 
\ , \qquad 
c_{\rm p}^{({\rm COL})}[q] = q (q-1) \alpha_{\rm p}^{({\rm XORSAT})}[k=q] \ .
\label{eq_param_col}
\end{equation}
The initiated reader will recognize the order parameter as the fraction of 
hard fields in the solution of the 1RSB equations at $m=1$ given 
in~\cite{long_col}; we shall come back on this point later on.

The determination of the threshold $\alpha_{\rm p}$ is slightly more
involved in the SAT problem. We have indeed a family of distributions
$q_n(h)$, $\hq_n(u)$ indexed by a real $h$, $u$; it it thus necessary 
to define for each of them an order parameter $\phi(h)$, $\hphi(u)$, as the 
fraction of infinite values of $n$ born by $q_n(h)$, $\hq_n(u)$.
The equivalent of Eq.~(\ref{eq_param_xor}) takes now a functional form 
easily derived from Eqs.~(\ref{eq_hq_sat},\ref{eq_q_sat}),
\begin{eqnarray}
\hphi(u) \hPP(u) &=& \int \prod_{i=1}^{k-1} d\P(h_i) 
\ \delta(u-f(h_1,\dots,h_{k-1})) 
\prod_{i=1}^{k-1} \frac{1-\tanh h_i}{2} \phi(h_i) \ , 
\label{eq_param_sat_htou}
\\
\phi(h) \P(h) &=& \sum_{l_+,l_-=0}^\infty p_{l_+,l_-}
\int \prod_{i=1}^{l_+} d\hPP(u_i^+)
\prod_{i=1}^{l_-} d\hPP(u_i^-) \ 
\delta\left(h -\sum_{i=1}^{l_+} u_i^+ + \sum_{i=1}^{l_-} u_i^- \right)
\left(1 - \prod_{i=1}^{l_-}(1-\hphi(u_i^-)) \right) \ . 
\label{eq_param_sat_utoh}
\end{eqnarray}
From the solution of these equations the order parameter of the average 
m.s.r.d. is obtained (see Eq.~(\ref{eq_sat_final})) as 
$\phi = \int d\P(h) (1-\tanh h) \phi(h)$. Again, $\phi$ is the fraction
of hard fields in the $m=1$ 1RSB equations of~\cite{long_sat}, this connection
shall be discussed further in Sec.~\ref{sec_msr_1rsb} and 
App.~\ref{sec_app_1rsb}. A solution of the functional
equation on $\phi(h)$ can be sought by a population dynamics algorithm:
the distribution $\P(h)$ being represented by a sample $\{h_i\}$, we
associate to each of them an estimation $\phi_i$ of $\phi(h_i)$ and
consider a population of couples $\{(h_i , \phi_i ) \}_{i=1}^{\cal N}$.
From this a new population $\{u_j,\hphi_j\}_{j={\cal N}+1}^{2{\cal N}}$
is generated according to Eq.~(\ref{eq_param_sat_htou}): for each element
of the new population $k-1$ indices $i_1,\dots,i_{k-1}$ are chosen uniformly
at random in $[1,{\cal N}]$ and the new couple $(u_j,\hphi_j)$ is computed as
\begin{equation}
(u_j,\hphi_j) = \left(f(h_{i_1},\dots,h_{i_{k-1}}), \prod_{m=1}^{k-1}
\frac{1-\tanh h_{i_m}}{2} \phi_{i_m}  \right) \ .
\end{equation}
In turns the sample $\{(h_i , \phi_i ) \}$ is generated
from the $\{u_i,\hphi_i\}$'s according to (\ref{eq_param_sat_utoh}), and
an estimation for the order parameter is computed as
\begin{equation}
\phi = \frac{1}{\cal N} \sum_{i=1}^{\cal N} (1 - \tanh h_i) \, \phi_i \ .
\end{equation}
These two steps are iterated a large number of times, starting with the
initial condition $\phi(h)=1$, i.e. $\phi_i=1$ for all elements of the initial
population. For small values of $\alpha$ the function
$\phi(h)$ converges to 0 upon these iterations, while for larger values
a non-trivial fixed point is reached. The numerical estimation of the 
threshold $\alpha_{\rm p}$ separating these two regimes, along with 
the deduced order parameter at the transition, are presented in 
Table~\ref{tab}. The precision on these numbers is rather low; indeed,
strong finite $\cal N$ effects make difficult a precise determination of the 
discontinuous disappearance of the non-trivial solution. Moreover
the numerical method becomes difficult for large values of $k$, hence
the limitation of the results presented to $k \in [3,6]$.
For $k=3$ $\phi_{\rm p}$ can also be successfully compared on the right part of 
Fig.~\ref{fig_results} with the plateau in the numerically determined $Q_n$.

\begin{table}
\begin{tabular}{| c || c | c || c | c || c | c | c | c || c | c | c | c | c | c |}
\hline
\multicolumn{1}{|c||}{}& \multicolumn{2}{|c||}{XORSAT} & \multicolumn{2}{|c||}{COL} & 
\multicolumn{4}{|c||}{XORSAT and COL} & \multicolumn{6}{|c|}{SAT} \\
\hline
$k,q$ & $\alpha_{\rm p}$ & $\phi_{\rm p}$ & $c_{\rm p}$ & $\phi_{\rm p}$ 
& $\lambda$ & $a$ & $b$ & $\nu$ & $\alpha_{\rm p}$ & $\phi_{\rm p}$ & $\lambda$ & $a$ & $b$ & $\nu$\\
\hline
\hline
3 & 0.818469 & 0.715332 & 4.910815 & 0.511700 & 0.397953 & 0.422096 & 1.221834 & 1.593787 &  4.40 & 0.74 & 0.55 & 0.38 & 0.90 & 1.87 \\
\hline
4 & 0.772280 & 0.851001 & 9.267358 & 0.616297 & 0.350174 & 0.433412 & 1.341647 & 1.526313 & 10.55 & 0.86 & 0.40 & 0.42 & 1.22 & 1.60 \\
\hline
5 & 0.701780 & 0.903350 & 14.035605 & 0.665924 & 0.320971 & 0.439997 & 1.421808 & 1.488035 & 21.22 & 0.91 & 0.33 & 0.44 & 1.40 & 1.50 \\
\hline
6 & 0.637081 & 0.930080 & 19.112434 & 0.695986 & 0.300707 & 0.444431 & 1.481191 & 1.462601 & 39.87 & 0.93 & 0.31 & 0.44 & 1.45 & 1.47 \\
\hline
7 & 0.581775 & 0.945975 & 24.434557 & 0.716600 & 0.285554 & 0.447677 & 1.527913 & 1.444121 & & & & & & \\
\hline
8 & 0.534997 & 0.956381 & 29.959848 & 0.731841 & 0.273649 & 0.450187 & 1.566174 & 1.429899 & & & & & & \\
\hline
9 & 0.495255 & 0.963661 & 35.658363 & 0.743697 & 0.263961 & 0.452205 & 1.598411 & 1.418505 & & & & & & \\
\hline
10 & 0.461197 & 0.969008 & 41.507763 & 0.753261 & 0.255868 & 0.453873 & 1.626162 & 1.409102 & & & & & & \\
\hline
\hline
\end{tabular}
\caption{Threshold, order parameter and critical exponents for the freezing 
transition in random tree ensembles.}
\label{tab}
\end{table}

\begin{figure}
\includegraphics[width=7cm]{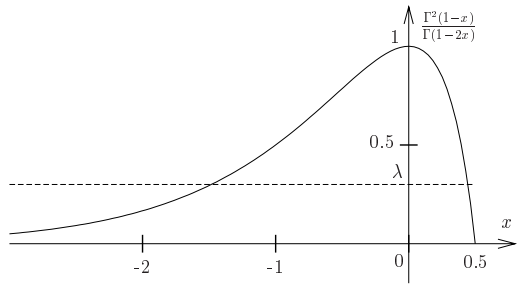}
\caption{The exponent $a$ (respectively $-b$) is the positive (respectively 
negative) root of the equation represented here, see 
Eq.~(\ref{eq_MCT}).} 
\label{fig_eq_MCT}
\end{figure}

The discontinuous character of the transition exhibited by the jump of the
order parameter should not hide the strong precursor effects, usually
associated to continuous transitions, present in the low connectivity phase.
The existence of a diverging scale of rearrangement sizes
is indeed obvious on Fig.~\ref{fig_results}. One can for instance define 
$n_\epsilon(\alpha)$ as the point where $Q_n$ crosses a threshold
$\epsilon$. This scale $n_\epsilon(\alpha)$ diverges at $\alpha_{\rm p}$
(as long as $0<\epsilon < \phi_{\rm p}$), in other words arbitrary large 
rearrangements are present with positive probability sufficiently close
to the transition. A detailed study of the XORSAT problem~\cite{MoSe}, 
drawing on 
a formal analogy with the mode-coupling theory of supercooled 
liquids~\cite{MCT}, revealed that the divergence of $n_\epsilon$ is algebraic, 
$n_\epsilon \sim (\alpha_{\rm p} - \alpha)^{- \nu}$. This exponent $\nu$ is
the solution of an universal type of relations,
\begin{equation}
\nu = \frac{1}{2a}+\frac{1}{2b} \ , \qquad 
\frac{\Gamma^2(1-a)}{\Gamma(1-2a)}= \frac{\Gamma^2(1+b)}{\Gamma(1+2b)} =
\lambda \ , 
\label{eq_MCT}
\end{equation}
where $\Gamma$ denotes Euler's special function (see Fig.~\ref{fig_eq_MCT}) and
$\lambda$ a $k$-dependent parameter in $[0,1]$. In fact $a$ and $b$ are
also critical exponents governing the asymptotic behavior of $Q_n$ around 
its plateau, see App.~\ref{sec_app} for details. The non-universal parameter 
$\lambda$ was found~\cite{MoSe} to be, in the XORSAT case,
\begin{equation}
\lambda^{({\rm XORSAT})} = 
\frac{k-2}{\alpha_{\rm p} k (k-1) \phi_{\rm p}^{k-1}} \ .
\label{eq_lambda_xorsat}
\end{equation}
Numerical values of this parameter and the associated exponents $a,b,\nu$
can be found in Table~\ref{tab}. Because of 
Eq.~(\ref{eq_traduction}) the exponents for the $q$-coloring are exactly the 
same as the one of $k$-XORSAT, provided one identifies $k$ and $q$. It will
be useful for future discussion to rewrite the parameter $\lambda$ under the
form
\begin{equation}
\lambda^{({\rm COL})} = (q-2)
\frac{1 - \phi_{\rm p}^{1/(q-1)}}{\phi_{\rm p}^{1/(q-1)}} \ .
\label{eq_lambda_col}
\end{equation}

The asymptotic behavior of the distribution $q_n$ for SAT could be a
priori more complicated, because of the underlying infinity of distributions
$q_n(h)$. We shall however argue in App.~\ref{sec_app} that the phenomenology
remains the same, in particular the exponents $a,b$ and $\nu$ are still given 
by Eq.~(\ref{eq_MCT}). The parameter $\lambda$ is now
\begin{equation}
\lambda^{({\rm SAT})} = 
\frac{2^k (k-2)}{\alpha_{\rm p} k (k-1) \phi_{\rm p}^{k-1}} \ ,
\label{eq_lambda_sat}
\end{equation}
the expression (\ref{eq_lambda_xorsat}) being only modified by a scale 
factor $2^k$ on the connectivity. 
The value of $\lambda$ can thus be determined from the numerical evaluation of 
$\alpha_{\rm p}$ and $\phi_{\rm p}$ explained above (see Table~\ref{tab} for
the results). The technical details of the analysis, along with numerical
evidence supporting it, can be found in App.~\ref{sec_app}.

\section{A digression about the reconstruction problem}
\label{sec_reconstruction}

It is instructive, and shall be useful for the discussion of the following
Section, to reconsider the freezing transition from a slightly different
perspective, namely the problem of tree reconstruction~\cite{Mossel}.
For simplicity we consider first the $q$-colorings of regular trees with
$L+1$ generations, where every vertex has degree $l+1$ (apart from the root 
which has degree $l$ and from the leaves of degree 1). The generation of an 
uniform proper coloring can be seen as a broadcasting process: the color of 
the root being chosen, each of its sons has a color uniformly chosen among 
the $q-1$ other ones, and this is propagated until the leaves of depth $L$ 
have been reached. 
In an information theoretic vision the color of the root is an information
transmitted through a noisy channel, the tree. The reconstruction problem
consists in infering the color of the root given the observation of the 
colors of the leaves, while the rest of the coloring is hidden to the
observer. Depending on the values of $(l,q)$ a correlation between the color
of the root and the one of the leaves survives or not the limit $L \to \infty$.
An optimal algorithm will be able to infer the value of the root from the
observation of the leaves with a probability of success larger than the 
one of a random uniform guess if and only if this correlation remains 
positive. In
this case the reconstruction problem is said to be solvable,
which can also be formulated as the non-extremality (or impurity) of the 
free-boundary Gibbs measure~\cite{Giorgii} on the infinite tree. On general
grounds one expects a critical value $l_{\rm d}(q)$ separating a solvable
regime for $l \ge l_{\rm d}(q)$ and an unsolvable one when $l < l_{\rm d}(q)$.
The values $l_{\rm d}(3)=5$, $l_{\rm d}(4)=8$, $l_{\rm d}(5)=13$ and 
$l_{\rm d}(6)=17$ have been conjectured in~\cite{MeMo}, along with rigorous
bounds $l_{\rm d}(3) \le 5$, $l_{\rm d}(4) \le 9$, $l_{\rm d}(5) \le 13$ and 
$l_{\rm d}(6) \le 17$. 

A very naive, suboptimal algorithm to perform this inference proceeds from 
the leaves towards the root, according to
the following rule: if the set of colors on the descendants of a vertex $i$
contains $q-1$ distinct elements in $[1,q]$, the remaining color is assigned
to the vertex $\s_i$. Otherwise it is assigned a neutral color, say white,
$\s_i=0$. It is easy to realize that at the end of the execution of this 
algorithm, starting from the
observation of the leaves of a proper coloring, the vertices in the interior
of the tree are either white or have been assigned the correct value they
had in the initial coloring. What is the probability $\phi_L$ (with respect
to the choice of the initial coloring) that the root is correctly reconstructed
in such a way? For this to be possible, $q-1$ distinct colors had to be
assigned to its sons in the initial coloring, and for each color at least one
of them had to be correctly reconstructed. $\phi_L$ can thus be
determined by recurrence according to $\phi_{L+1} = V(\phi_L)$, with
\begin{equation}
V(\phi) = \frac{1}{(q-1)^l} 
\sum_{\substack{l_1,\dots,l_{q-1} \\ \sum l_i =l }} 
\frac{l!}{l_1! \dots l_{q-1}! } \prod_{i=1}^{q-1} \left( 1 - (1-\phi)^{l_i} 
\right) = \sum_{p=0}^{q-1} \binom{q-1}{p} (-1)^p 
\left( 1 - \frac{p}{q-1} \phi \right)^l \ ,
\label{eq_prob_reconstruction}
\end{equation}
and $\phi_{L=0}=1$. The limit of $\phi_L$ for large $L$ is the largest
solution of the fixed-point equation $\phi = V(\phi)$ on the interval
$[0,1]$. 
Depending on the values of $q$ and $l$ this limit is either zero (for instance
$V$ vanishes identically if $l < q-1$ as there are not enough descendants for 
the root to be fully constrained) or strictly positive. 
By numerical inspection
of Eq.~(\ref{eq_prob_reconstruction}) we found the latter case to happen
when $l \ge l_{\rm p}(q)$, with $l_{\rm p}(3)=5$, $l_{\rm p}(4)=9$, 
$l_{\rm p}(5)=14$ and $l_{\rm p}(6)=19$. This means that the algorithm
has a positive probability of guessing correctly the root from the observation 
of arbitrarily distant leaves when $l \ge l_{\rm p}(q)$, whereas it is doomed
to fail if $l < l_{\rm p}(q)$. 
The reasoning presented here is essentially a constructive proof of the bound 
$l_{\rm d}(q) \le l_{\rm p}(q)$, weaker yet conceptually much simpler than the
rigorous bound of~\cite{MeMo}. Let us underline that such a reconstruction
procedure is far from optimal; we only retain the
information given by a drastic event, when the color of a vertex is 
unambiguously determined by its descendents, and discard the cases where
one color is only more probable than the others.

This naive reconstruction algorithm is in a sense dual with the main subject 
of the paper: it correctly infers the color of the root if and only in all
proper colorings with the observed assignment of the leaves the root takes
always the same color. In other words in all rearrangements (not necessarily
of minimum size) for the root starting from the initial coloring at least one
site on the boundary of the tree has to be rearranged. This can be determined
using the recursion relation on the sizes of the minimal rearrangements
(for instance (\ref{eq_n_itoj_col},\ref{eq_n_i_col}) in the case of the 
coloring) with a different boundary condition, $n_{i \to j}=\infty$ when $i$ 
is a leaf of the tree. The value $n_i$ computed with this boundary condition
will be infinite if there are no rearrangements of the root which can avoid
rearranging the leaves (the algorithm is successful), finite otherwise
(the root is white at the end of the algorithm). This difference in the
boundary condition ($n_{i \to j}=\infty$ vs $1$) is irrelevant in the large
$L$ limit: m.s.r. of finite size have supports of finite depth, hence are not 
affected by the boundary when $L$ gets larger than this depth, while m.s.r.
of sizes growing with $L$ are correctly assigned their formal infinite size
in this way. 

To summarize the connection between this Section and the rest of the paper,
the constraints that imply large rearrangements are
precisely the information exploited in the naive reconstruction procedure. The
probability of success of the algorithm on arbitrarily large trees can thus be
identified with the order parameter of the freezing transition introduced in
the previous Section. This identification holds for generic CSPs on random
trees, provided one averages the success probability of the naive
reconstruction over the ensemble of trees. 
Another suggestive perspective on the problem is given in terms of percolation.
The order parameter can indeed be viewed as the probability of percolation of
the support of the rearrangement from the root to an infinitely distant
boundary. In the case of XORSAT this percolation is purely geometrical and
corresponds to the existence of an infinite subtree where all variable nodes
have degree greater than two. For COL and SAT the object which percolates is
subtler: the rearrangements depend both on the geometry of the factor
graph and on its initial solution.

\section{From random trees to random graphs}
\label{sec_rgraph}

\subsection{Local and global aspects of the cavity method}
\label{sec_lwc}

We turn now to the more delicate issue of the validity of the results
derived in the random tree ensembles for the original random graphs.
As mentioned in Sec.~\ref{sec_def} the latter have a local tree structure,
with high probability in the thermodynamic limit. The point thus amounts
to giving a description of the boundary condition induced by the rest of the
factor graph. We shall handle this problem in the framework of the cavity
method~\cite{beyond,book_Talagrand} for sparse random graphs~\cite{MePa} 
(see also~\cite{MeMo} for a related discussion) that we briefly survey
below.

Consider a $G(N,M)$\footnote{random hypergraphs with arbitrary degree 
distributions can be studied similarly.} random factor graph $F$ with $N$ 
variable nodes and 
$M=\alpha N$ constraint nodes of degree $k$, the associated random measure on 
$\X^N$,
\begin{equation}
\mu(\us ) = \frac{1}{Z} \prod_{a=1}^M \psi_a(\us_a) \ ,
\label{eq_rgraph_law}
\end{equation}
and suppose the weights $\psi_a$ are i.i.d. positive random functions
on $\X^k$ (not necessarily $\{0,1\}$ valued).

Two kind of intertwined properties of the model can be investigated:
thermodynamic (global) ones, with the characterization of the random variable 
$Z$, and local ones, concerning the behavior of the measure $\mu$ itself. 
Because of the self-averaging properties of $\ln Z$ for large graphs
the central thermodynamic quantity is the quenched free-entropy density,
\begin{equation}
\Phi = \lim_{N \to \infty} \frac{1}{N} \E \ln Z \ .
\end{equation}
The latter aspect of the problem, which is the important one for
our present concerns, can be formulated as follows. Call $F_L$ the sub-factor
graph induced in $F$ by the variable nodes at a graph distance smaller than $L$
from an arbitrary root $i$, and $\us_L$ the configuration of these variable
nodes. As we are interested in the thermodynamic limit with $L$ finite we can
assume without harm that $F_L$ is a tree. The marginalization of 
Eq.~(\ref{eq_rgraph_law}) leads to a law $\mu_L$ for $\us_L$; it can be 
seen as a random measure, conditioning $F$ on a given realization of $F_L$,
because of the choices in $F \setminus F_L$. At this point a question arises
naturally: what is the (weak) limit of $\mu_L$ when the 
thermodynamic limit is taken?

The cavity method provides a series of possible answers to this question,
and an heuristic to choose the right one. Let us introduce some notations:
we denote $b$ the number of sites in the boundary $B$ made of the sites
at distance exactly $L$ from $i$, $B = \{i_1, \dots, i_b \}$, and
define a measure on $\us_L$ with external fields $\eta_j$ 
(probability measures on $\X$) acting on this boundary:
\begin{equation}
\mu^{(0)}(\us_L ; \eta_1,\dots,\eta_b) = 
\frac{1}{Z_0(\eta_1,\dots,\eta_b)} \prod_{a \in F_L} \psi_a(\us_a )
\prod_{j=1}^b \eta_j(\s_{i_j}) \ ,
\label{eq_def_mu0}
\end{equation}
where $Z_0$ ensures the normalization of the law.

The statement of the simplest (so-called Replica-Symmetric, RS) situation
described by the cavity method is
\begin{equation}
\mu_L(\cdot ) \tod \mu^{(0)}(\cdot ; \eta_1,\dots,\eta_b ) \ ,
\label{eq_statement_RS}
\end{equation}
where the $\eta_i$ are i.i.d. from a distribution $\P_{(0)}$. Roughly 
speaking, this is true when $\mu$ is a (finite size) pure state, so that
the effect of $F \setminus F_L$ on the boundary variables can be factorized.
In more complicated situations there is a large number of pure states on
which the Gibbs measure has to be decomposed for this factorization to
be possible\footnote{We skip the intermediate case of a finite number of pure
  states; for instance the low temperature phase of an Ising ferromagnet
  should be described by the superposition of the two $\mu^{(0)}$ of positive
  and negative magnetization.}. 
We shall thus introduce a new measure on $\us_L$ as a weighted
superposition of different $\mu^{(0)}$,
\begin{equation}
\mu^{(1)}(\us_L ; P^{(1)}_1,\dots,P^{(1)}_b,m) =
\frac{1}{Z_1[P^{(1)}_1,\dots,P^{(1)}_b;m]} \int \prod_{i=1}^b
dP^{(1)}_i(\eta_i) \ \mu^{(0)}(\us_L ; \eta_1,\dots,\eta_b ) 
\ Z_0(\eta_1,\dots,\eta_b)^m \ .
\label{eq_def_mu1}
\end{equation}
In this definition $m \in [0,1]$ is known as the Parisi breaking parameter, 
the $P_i^{(1)}$'s are
distributions of fields, and again $Z_1$ is a normalization.
The hypothesis of the cavity method at the level of one step of 
Replica-Symmetry Breaking (1RSB) reads
\begin{equation}
\mu_L(\cdot ) \tod \mu^{(1)}(\cdot ; P^{(1)}_1,\dots,P^{(1)}_b,m) \ ,
\end{equation}
where the $P^{(1)}_i$ are i.i.d. from a distribution $\P_{(1)}$. 
In some cases the 1RSB description coincides with the RS one, for instance
whenever the $P^{(1)}_i$ in the support of $\P_{(1)}$ are concentrated on a
single value of the field (in the following we shall call this a trivial 1RSB 
solution). A less obvious reduction happens when the parameter
$m$ is equal to 1: from (\ref{eq_def_mu0},\ref{eq_def_mu1}) one realizes that 
in this case $\mu^{(1)}$ is indistiguinshable from $\mu^{(0)}$ with properly 
averaged values of the external fields, more precisely
\begin{equation}
\mu^{(1)}(\us_L ; P^{(1)}_1,\dots,P^{(1)}_b,m=1) =
\mu^{(0)}(\us_L ; \oeta_1,\dots,\oeta_b) \quad \text{with} \quad
\oeta_i = \int dP_i^{(1)}(\eta) \ \eta \ .
\label{eq_equi_RS_1RSBm1}
\end{equation}

This 1RSB formalism can be promoted to an arbitrary level of symmetry 
breaking by a recursive construction. Let us call $\M_0$ the set of 
probability laws on $\X$, and define by recurrence $\M_{K+1}$ as the set of
probability laws on $\M_K$. The measure $\mu^{(K)}$ with $K$ steps
of replica symmetry breaking is parameterized by $K$ reals
$0\le m_1 \le \dots \le m_K \le 1$ and $b$ elements $P_i^{(K)}$ of $\M_K$, 
and can be expressed recursively as
\begin{multline}
\mu^{(K+1)}(\us_L ; \{P^{(K+1)}_i\}_{i=1}^b,m_1,\dots,m_{K+1}) =
\frac{1}{Z_{K+1}[\{P^{(K+1)}_i\};m_1,\dots,m_{K+1}]} \\
\int \prod_{i=1}^b dP^{(K+1)}_i(P^{(K)}_i) 
\ \mu^{(K)}(\us_L ; \{P^{(K)}_i\}, m_2,\dots,m_{K+1}) 
\ Z_K(\{P^{(K)}_i\}, m_2,\dots,m_{K+1})^{m_1/m_2} \ .
\end{multline}
The $K$-RSB assumption of the cavity method reads
\begin{equation}
\mu_L(\cdot ) \tod \mu^{(K)}(\cdot ; P^{(K)}_1,\dots,P^{(K)}_b,m_1,\dots,m_K) 
\ ,
\label{eq_cavity_K}
\end{equation}
with the $P^{(K)}_i$ i.i.d. from $\P_{(K)}$, a given element of $\M_{K+1}$.
Eventually the limit of an infinite number of steps of replica symmetry 
breaking ($K \to \infty$) can be formally taken. Note that, as discussed
in the 1RSB case, $\mu^{(K)}$ incorporates as special cases (when the 
distributions concentrates on a single value, or when the $m_i$'s are 
degenerate) all possible descriptions at a smaller level of RSB.

We face now the problem of choosing, among all these possible assumptions,
which is the correct one. A first condition on the allowed values of 
$\P_{(K)}$ arises from a simple consistency requirement. $\mu_L$ can indeed be 
obtained in two ways: from a direct application of the statement 
(\ref{eq_cavity_K}), or by considering a larger neighborhood of depth $L'>L$ 
and making a partial marginalization of $\mu_{L'}$. As $F_{L'} \setminus F_L$ 
is distributed according to a Galton Watson branching process, the 
consistency of these various ways of obtaining $\mu_L$ induces conditions
restricting the possible values of $\P_{(K)}$. At the RS ($K=0$) level
this is nothing but the stationarity property stated in Eq.~(\ref{eq_RS}). 
The heuristic for the choice of $K$ and the values of the breaking parameters
$m_i$ arises from the global aspect of the cavity method, 
namely the computation of
the typical value of the free-entropy density $\Phi$. More precisely, for each
level of the RSB hierarchy there is a functional 
$\Phi_{(K)}[\P_{(K)},m_1,\dots,m_K]$ whose minimum is taken as an
estimation of $\Phi$. The bounds $\Phi \le \Phi_{(K)}$ have indeed
been rigorously proven in some cases~\cite{Guerra,FrLe,PaTa}, 
and are expected to hold with a certain generality. The best estimation
of $\Phi$, which is presumably exact in mean-field models (this has been
proven in one case~\cite{Ta_SK}), should thus be sought through the 
minimization
of $\Phi_{(K)}$ in the formal $K\to \infty$ limit which encompasses all
possible levels of RSB. The limit of $\mu_L$ is expected to be described
by the set of parameters achieving this minimum (note that the extremization
of $\Phi_{(K)}$ with respect to $\P_{(K)}$ corresponds to the consistency
requirement explained above). This minimization is obviously a formidable 
task which seems out of reach in its full generality for models on
sparse random graphs. There are however partial arguments which can be used
to assess the validity of the simplest RS and 1RSB hypothesis. The decay of 
point-to-set correlations at large distance
(in other words the purity of the Gibbs measure, or the non reconstructibility
of the value of a spin from the observation of distant sites) is indeed 
related to the absence of a non-trivial solution of the 1RSB consistency
equations at $m=1$~\cite{MeMo}, and suggests the RS hypothesis to be correct.
A test of the plausibility of the 1RSB description is usually performed via a
local stability analysis~\cite{stab}: one checks in this way the absence of a 
non-trivial solution of the 2RSB consistency equations in the vicinity of a 
1RSB solution $\P_{(1)}$.

Let us finally underline the deep connection between these issues and
the local weak convergence method developed by Aldous 
(see~\cite{Aldous,Aldous2} for reviews) on related optimization problems.
Recently the above stated local properties of the RS cavity method were
rigorously proven in some discrete models (cf. for 
instance~\cite{BaGa,MoSh,BaNa}), under a priori non-optimal conditions
(worst-case vs typical decay of correlations, i.e. uniqueness vs extremality 
conditions~\cite{letter}).

\subsection{Minimal size rearrangements in the random graph ensembles}

We shall now reconsider the computations of the m.s.r.d. performed in the
random tree ensembles in the light of the above presented cavity method.
It should be clear that the thermodynamic limit ($N \to \infty$) of the 
average distribution $q_n$ defined in Eq.~(\ref{eq_def_qn_rg}) for the 
original random graph ensembles coincide with the infinite $L$ limit of their 
tree counterpart whenever the RS assumption stated in (\ref{eq_statement_RS}) 
is valid. The probability measure on the initial configuration we used
for the computation of the m.s.r. in finite tree formulae (cf. 
Eq.~(\ref{eq_def_mu})) corresponds indeed to the limit measure $\mu^{(0)}$ 
on the finite neighborhood of the random graphs. The validity of this RS
scenario depends on the particular model and on the value of the connectivity
parameter $\alpha$ ($c$ for coloring).

In the case of XORSAT~\cite{xor1,xor2} the local properties of the uniform
measure over the set of solutions are well described by the RS assumption
upto the satisfiability threshold $\alpha_{\rm s}$, for all values of $k$.
In consequence the computation of $q_n$ performed in the random tree ensemble 
extends to random graphs throughout
the satisfiable phase $\alpha \le \alpha_{\rm s}$, the threshold for 
the freezing transition in random graphs ($\alpha_{\rm f}$) and in random 
trees ($\alpha_{\rm p}$) are equal, and the exponents governing the divergence
of the m.s.r. in the limit $\alpha \to \alpha_{\rm f}^-$ are correctly
obtained from (\ref{eq_lambda_xorsat}).
In fact $\alpha_{\rm p}$ corresponds also to
the clustering transition $\alpha_{\rm d}$ due to the appearance of an 
extensive 2-core: a rearrangement for a variable in the 2-core (more precisely
in the backbone~\cite{xor1,xor2}) is necessarily of extensive size. 
In agreement with this correspondence, the order parameter of the freezing
transition solution of (\ref{eq_param_xor}) is precisely the fraction of
vertices in the backbone.

The picture of the satisfiable phase of random $k$-SAT and $q$-COL advocated
in~\cite{letter,long_sat,long_col} is richer. Let us first describe it on
the example of SAT. At low values of the 
connectivities, $\alpha < \alpha_{\rm d}(k)$, one expects a plain RS 
description to hold. The clustering transition $\alpha_{\rm d}(k)$ corresponds
to the appearance of long-range point-to-set correlations, in other
words to a non-trivial solution of the 1RSB equations with $m=1$. In an 
intermediate regime $[\alpha_{\rm d}(k),\alpha_{\rm c}(k)]$ the thermodynamics
of the system is described by a 1RSB scenario with $m=1$, the dominant clusters
of solutions are exponentially numerous (their complexity is strictly 
positive)\footnote{The case $k=3$ is special from this point of view,
one finds indeed $\alpha_{\rm d}(3)=\alpha_{\rm c}(3)$ and
no intermediate phase with an exponential number of relevant clusters.}. 
At $\alpha_{\rm c}(k)$ a condensation phenomenon occurs, the
degeneracy of the thermodynamically relevant clusters becomes sub-exponential, 
and the 1RSB breaking parameter $m$ decreases from 1 to 0 as $\alpha$ 
increases
from $\alpha_{\rm c}(k)$ to the satisfiability threshold $\alpha_{\rm s}(k)$.
Higher levels of RSB might be necessary to describe the condensated regime 
$[\alpha_{\rm c}(k),\alpha_{\rm s}(k)]$; we shall in the following make the
hypothesis (partly supported by~\cite{stab2}) that this is not the case for 
$\alpha \le \alpha_{\rm c}(k)$. Because of the equivalence, for the local
properties of the measure, of an RS description and a 1RSB with $m=1$ 
(cf. Eq.~(\ref{eq_equi_RS_1RSBm1})), we thus expect the computation of the
minimal size rearrangements performed on the tree to be correct for random SAT
formulae with $\alpha \le \alpha_{\rm c}(k)$. For the sake of readability we
reproduce in Table~\ref{tab_thr_rgraphs} the values of $\alpha_{\rm d}(k)$
and $\alpha_{\rm c}(k)$ obtained in~\cite{letter,long_sat}, along with the 
satisfiability threshold $\alpha_{\rm s}(k)$ of~\cite{MeMeZe}.

Depending on the values of $k$ the freezing threshold $\alpha_{\rm p}(k)$
for the random tree ensemble is, or not, smaller than the
condensation one. For $k\in [3,5]$ one finds
$\alpha_{\rm p}(k) > \alpha_{\rm c}(k)$: for these values of $k$ 
the computation in the tree ensemble does not allow
the determination of the freezing threshold of the original ensembles 
$\alpha_{\rm f}(k)$ (at this point we can just say that 
$\alpha_{\rm f}(k) > \alpha_{\rm c}(k)$). For $k=6$ the situation is reversed, 
$\alpha_{\rm p}(6) < \alpha_{\rm c}(6) $, we thus conclude that 
$\alpha_{\rm f} (6)= \alpha_{\rm p} (6)$, and that the exponents $a,b,\nu$
describing the precursors of the freezing transition can be safely
computed from (\ref{eq_lambda_sat}). We expect the ordering of the various
thresholds, and hence the validity of the conclusions just stated for $k=6$,
to remain the same for all greater values of $k$. This is corroborated
by an analysis of the large $k$ limit presented in App.~\ref{sec_app_largek}:
the asymptotic behavior of $\alpha_{\rm p}(k)$ is much smaller than the
one of $\alpha_{\rm c}(k)$~\cite{letter,long_sat},
\begin{equation}
\alpha_{\rm f}(k)=\alpha_{\rm p}(k) = \frac{2^k}{k} ( \ln k + O(\ln \ln k))
\ll
\alpha_{\rm c}(k) = 2^k \ln 2 - O(1) \ .
\label{eq_sat_largek}
\end{equation}
In fact the SAT problem in the limit of large $k$ becomes similar to the 
XORSAT problem: the threshold $\alpha_{\rm f}(k) = \alpha_{\rm p}(k)$ is
equivalent to $2^k$ times the corresponding value for XORSAT, the order
parameter at the transitions are equivalent in both problems, hence the
parameter $\lambda$ governing the critical exponents becomes the same in
the large $k$ limit. Moreover from the results of~\cite{letter,long_sat}
on the behavior of the clustering threshold one realizes that the regime 
$[\alpha_{\rm d}(k),\alpha_{\rm f}(k)]$ where clusters are present yet do
not have frozen variables is of vanishing width in this limit.

The picture of the satisfiable regime for the $q$-coloring of random graphs
presented in~\cite{long_col} is essentially the same as the one of SAT we just
described. The dynamical, condensation and satisfiability thresholds
obtained in~\cite{long_col} are recalled in Table~\ref{tab_thr_rgraphs}
(the last two are denoted $c_{\rm g}(q)$ and $c_{\rm q}(q)$ 
in~\cite{long_col}). As argued above the computation performed in the
random tree ensemble should be correct for Poissonian random graphs of
mean connectivity $c \le c_{\rm c}(q)$; for $q \in [3,8]$ this regime
does not include the tree freezing transition $c_{\rm p}(q)$ (called
$c_{\rm r}(m=1)$ in~\cite{long_col}). Conversely for $q \ge 9$ we
have $c_{\rm f}(q) =c_{\rm p}(q)$, which is given exactly by $q(q-1)$ times 
the threshold of XORSAT (recall the formal equivalence between XORSAT and the 
free boundary COL problem stated in Eq.~(\ref{eq_traduction})), and the 
exponents $a$, $b$, $\nu$ are the same as in XORSAT (identifying $q$ and $k$).
This ordering of the thresholds is confirmed by the analysis at large $q$,
\begin{equation}
c_{\rm f}(q) = q (\ln q + O(\ln \ln q)) \ll 
c_{\rm c}(q) = 2 q \ln q - O(\ln q) \ ,
\label{eq_col_largeq}
\end{equation}
the behavior of $c_{\rm f}(q)$ being justified in App.~\ref{sec_app_largek} 
while the one of the condensation threshold was given in~\cite{long_col}.

\begin{table}
\begin{tabular}{| c || c | c | c | c | c || c | c | c | c | c |}
\hline
\multicolumn{1}{|c||}{}& \multicolumn{5}{|c||}{COL} & 
\multicolumn{5}{|c|}{SAT}\\
\hline
$k,q$ & $c_{\rm d}$ & $c_{\rm f}$ & $c_{\rm p}$ & $c_{\rm c}$ & $c_{\rm s}$ &
$\alpha_{\rm d}$ & $\alpha_{\rm f}$ & $\alpha_{\rm p}$ & 
$\alpha_{\rm c}$ & $\alpha_{\rm s}$ \\
\hline
\hline
3 & 4     & 4.6 & 4.911 & 4 & 4.68 & 3.86 &  & 4.40 & 3.86 & 4.267 \\
\hline
4 & 8.35  & 8.8 & 9.267 & 8.4 & 8.90 & 9.38 &  & 10.55 & 9.547 & 9.931 \\
\hline
5 & 12.83 & 13.5 & 14.036 & 13.2 & 13.67 & 19.16 &  & 21.22 & 20.80 & 21.117 \\
\hline
6 & 17.64 & 18.6 & 19.112 & 18.4& 18.88& 36.53 & \multicolumn{2}{|c|}{39.87} & 43.08 & 43.37 \\
\hline
7 & 22.70 & 24.1 & 24.435 & 24.0 & 24.45 & & \multicolumn{2}{|c|}{} & & \\
\hline
8 & 27.95 & 29.93 & 29.960 & 29.90 & 30.33 & &\multicolumn{2}{|c|}{}  & & \\
\hline
9 & 33.45 & \multicolumn{2}{|c|}{35.658} & 36.0 & 36.49 & & \multicolumn{2}{|c|}{} & & \\
\hline
10 & 39.0 & \multicolumn{2}{|c|}{41.508} & 42.5 & 42.9 & & \multicolumn{2}{|c|}{} & & \\
\hline
\hline
\end{tabular}
\caption{Thresholds for the original random ensembles. The COL values
are from~\cite{long_col} and~\cite{thresh_col} for the satisfiability 
threshold $c_{\rm s}$, the SAT ones from~\cite{letter,long_sat} 
and~\cite{MeMeZe} for $\alpha_{\rm s}$. For $q\in[3,8]$ the freezing threshold
of~\cite{long_col} is computed at the 1RSB level.}
\label{tab_thr_rgraphs}
\end{table}

\subsection{Dealing with RSB}
\label{sec_msr_1rsb}

We have thus reached the frustrating conclusion that the computations
performed upto now were not able to determine the average m.s.r.d. in
the condensated phase of SAT and COL, and in particular for $k\in [3,5]$, 
$q\in [3,8]$, to locate the freezing transition 
and describe its critical behavior. The presentation of the
cavity method of Sec.~\ref{sec_lwc} indicates clearly what has to be done
to remedy this insufficiency: one should reproduce the computations
of the m.s.r.d. on finite trees, taking for the probability law on the
initial configurations $\mu^{(K)}$ instead of the $\mu^{(0)}$ we initially
considered. This generalized computation can in fact be performed in a similar
way, at the price of some technical complications, 
and is sketched for the $K=1$ level of
replica symmetry breaking in Appendix~\ref{sec_app_1rsb}. The resulting 
equations become rather difficult to solve and we leave the complete
determination of the distribution $q_n$ as an open problem. One can however
draw some general observations that we want to underline here. The order 
parameter of the freezing transition, i.e. the fraction of rearrangements of
diverging size, corresponds to the probability (over the pure states 
distribution) of a variable being acted on by an hard field which constrains
it to a single value. This was found above in the three CSP we considered
when the freezing transition happens in a 1RSB phase with $m=1$, and will
be shown in App.~\ref{sec_app_1rsb} to hold in non trivial situations with 
$m<1$. This should remain true for any CSP and any further level of RSB.
Another universality statement concerns the critical behavior of the
distribution $q_n$ around the freezing transition $\alpha_{\rm f}$.
The phenomenology described by the exponents $a$, $b$, $\nu$ can indeed be
argued to persist even when $\alpha_{\rm f}$ belongs to the condensated
regime $[\alpha_{\rm c},\alpha_{\rm s}]$. Moreover the parameter $\lambda$
fixing the value of the exponents can be expressed from the standard RSB 
computation.
The reader will find in App.~\ref{sec_app_1rsb} the technical details leading 
to this conclusion for SAT and COL at the 1RSB level, which is also expected
to hold for other CSPs and higher levels of RSB.

\section{Conclusions and perspectives}
\label{sec_conclu}

One of the main themes of the paper was the distinction that has
to be made between the clustering and freezing transitions. These can 
coincide in sufficiently symmetric problems like XORSAT, yet in general 
the solution space gets clustered without variables taking the same
value in all elements of the clusters. A definition of the clustering
threshold $\alpha_{\rm d}$ was put forward in~\cite{letter} as the smallest 
connectivity such that the long-range point-to-set correlation
\begin{equation}
\lim_{L \to \infty} \lim_{N \to \infty} \E \sum_{\us_{\partial L}} 
\mu(\us_{\partial L})
\sum_{\s_i} | \mu(\s_i | \us_{\partial L}) - \mu(\s_i) | 
\end{equation}
remains positive, where $i$ is an arbitrary variable node and 
$\us_{\partial L}$ the
configuration of the nodes at graph distance exactly $L$ from $i$.
A similar definition of the freezing transition $\alpha_{\rm f}$ can be
given in terms of the stronger notion of correlation
\begin{equation}
\lim_{L \to \infty} \lim_{N \to \infty} \E \sum_{\us_{\partial L}} 
\mu(\us_{\partial L})
\sum_{\s_i} \I( \mu(\s_i | \us_{\partial L}) = 1  ) \ ,
\end{equation}
hence $\alpha_{\rm f} \ge \alpha_{\rm d} $. The sub-optimality of the naive 
reconstruction algorithm given in Sec.~\ref{sec_reconstruction} should
clarify why this inequality is in general strict.

In this paper we concentrated on the rearrangements of finite sizes in the
thermodynamic limit, i.e. we computed the limit $N \to \infty$ 
(or $L\to \infty$ in tree ensembles) of the distributions
$q_n$ at a fixed value of the sizes $n$. 
The percolating rearrangements thus appeared as formally 
infinite values of $n$ which had to be included to ensure the normalization
of the limiting $q_n$. It should be an interesting research problem to 
describe more precisely these diverging size rearrangements by taking a scaling
limit of $q_n$, letting $n$ grows with $N$. The leading order is expected
to be linear in $N$, as are the minimal Hamming distances between clusters
studied for instance in~\cite{dist_XOR}.

The divergence of the minimal size of rearrangements can be viewed as a
percolation phenomenon of their supports. In the case of XORSAT this is
nothing but the classical 2-core percolation of random hypergraphs; for
general CSP, in particular SAT and COL, the percolating structure is defined
in two steps, the factor graph being equipped with a measure on the set of
initial configurations. The universality of their critical behavior
described by the exponents $a$, $b$, $\nu$ and the relations (\ref{eq_MCT})
between them is shared by other similar problems, for instance
rigidity~\cite{rigidity} and $q$-core~\cite{kcore} percolation when defined
on Bethe lattices. The latter problem is strongly related to kinetically
constrained models~\cite{Sellitto}, for which minimal size rearrangements can
be also computed and have the same critical behavior~\cite{nostro_KCM}.

The recursion relations (\ref{eq_n_atoi_gene}) could form the basis of new
investigations on the structure of a single formula, following the line
of research pioneered in~\cite{SP_science}. Though there is no guarantee of 
convergence in the presence of cycles in the factor graph, they can be turned
into an heuristic message passing algorithm that will provide informations
on a solution of a given instance of CSP. This solution should be found by
an independent solver algorithm, or, as was proposed in~\cite{Jorge}, in
an incremental way. Starting from an empty formula and an arbitrary assignment
of the variables constraints are introduced one by one. Whenever the new
constraint is violated by the current assignment one rearranges it; 
in~\cite{Jorge} this step was performed by a local search algorithm, that 
could be replaced by the single sample m.s.r. message passing heuristic.

The study of the rearrangements of XORSAT performed in~\cite{MoSe} addressed
further issues left apart in the present work. 
One was the characterization of the
geometrical properties of the m.s.r., through the distribution of their
average depths and a measure of their cooperativity by a geometrical 
susceptibility. We expect some of these geometrical results to extend 
from XORSAT to arbitrary CSPs,
in particular the value of the critical exponents $\zeta=1/2$, $\eta=1$ 
(see~\cite{MoSe} for their definitions). Another aspect should on the 
contrary be much more problem dependent, namely the structure of the energy
barriers between rearranged configurations. Given a pair of satisfying
assignments $\us,\ut$ one can define the set of paths in the configuration 
space which leads from one to the other by modifying one variable at a time, 
each variable being modified at most once. The barrier between 
$\us$ and $\ut$ can be defined as the minimum over this set of paths of the
maximum along the path of the number of violated constraints. One can then
study the rearrangements which modify a given variable $i$ and achieves a
minimal value of the barrier between the initial and final configurations.
The structure of XORSAT is such that minimal barrier and minimal size
rearrangements do not coincide, and that energy barriers are always strictly
positive (unless the variable appears in no constraint, otherwise flipping
a variable always makes at least one constraint unsatisfied). On the contrary
for SAT a finite size rearrangement can always be performed remaining in the
set of satisfying configurations: one just has to flip the variables in
decreasing order with respect to the distance from the root of the 
rearrangement.

Let us finally mention that the general formalism can be applied to several
CSPs besides the three examples on which we concentrated. For instance the
bicoloring of random hypergraphs~\cite{bicoloring}, which admits a stationary
free boundary, is easily seen to have a freezing transition in random tree
ensembles with branching ratio
\begin{equation}
\alpha_{\rm p}^{\rm (BICOL)}(k) = 
\left( 2^{k-1} -1 \right) \alpha_{\rm p}^{\rm (XORSAT)}(k) \ . 
\end{equation}

\acknowledgments
I warmly thank
Florent Krzakala,
Andrea Montanari,
Federico Ricci-Tersenghi and
Lenka Zdeborov\'a for a fruitful collaboration, in particular A.~M. 
with whom some of the techniques were developed in~\cite{MoSe} and for
enlightening discussions on the issue of Sec.~\ref{sec_lwc}, and Jorge
Kurchan for interesting exchanges about~\cite{Jorge}.

The work was partially supported by EVERGROW, integrated project No. 1935 
in the
complex systems initiative of the Future and Emerging Technologies directorate 
of the IST Priority, EU Sixth Framework.

\appendix

\section{Critical behavior around the freezing transition}
\label{sec_app}

\subsection{XORSAT}

In this appendix we shall give some details on the asymptotic behavior
of the average m.s.r.d. in the neighborhood of the freezing transition
in the random tree ensembles. The case of XORSAT was treated 
in~\cite{MoSe}, the main interest will thus be in the extension of these
results to the SAT problem. For the sake of clarity we first recall briefly 
some of the key points of App.~C in~\cite{MoSe}.

Let us define the generating functions of $q_n$ and $\hq_n$ as
\begin{equation}
R(x) = \sum_{n=1}^\infty q_n x^n \ , \qquad \qquad 
\hR(x) = \sum_{n=1}^\infty \hq_n x^n \ .
\end{equation}
The equations~(\ref{eq_q_XORSAT},\ref{eq_hq_XORSAT}) can be rewritten as
\begin{eqnarray}
\hQ_n &=& Q_n^{k-1} \ , \label{eq_app_hQ} \\
R(x) &=& x \exp [- \alpha k + \alpha k \hR(x) ] \label{eq_app_R} \ .
\end{eqnarray}
The order parameter $\phi = \lim_{n \to \infty} Q_n$ can also be expressed
as $R(x=1)$; the equation determining $\phi$ is formally written
as $\phi=V(\phi,\alpha)$ with $V(\phi,\alpha)= 1-\exp [ -\alpha k \phi^{k-1}]$.
At the transition point $(\alpha_{\rm p},\phi_{\rm p})$ we have 
$\partial_\phi V = 1$: the two curves become tangent at this point.
More explicitly,
\begin{eqnarray}
\phi_{\rm p} &=& 1 - \exp [ -\alpha_{\rm p} k \phi_{\rm p}^{k-1} ] \ , 
\label{eq_app_order} \\
1 &=& \alpha_{\rm p} k(k-1) \phi_{\rm p}^{k-2} 
\exp [ -\alpha_{\rm p} k \phi_{\rm p}^{k-1} ] \ .
\label{eq_app_derivative}
\end{eqnarray}

Consider first the large $n$ regime right at the transition 
($\alpha=\alpha_{\rm p}$), and assume that the decay of $Q_n$ towards
the plateau $\phi_{\rm p}$ is algebraic,
$Q_n \sim \phi_{\rm p} + A \ n^{-a}$, with $A$ a positive constant and $a$ 
a positive exponent. Expanding Eq.~(\ref{eq_app_hQ}) with this ansatz, 
we obtain
\begin{equation}
\hQ_n \sim \phi_{\rm p}^{k-1} + (k-1) \phi_{\rm p}^{k-2} A \ n^{-a}
+ \frac{(k-1)(k-2)}{2} \phi_{\rm p}^{k-3} A^2 \ n^{-2a} \ .
\end{equation}
The properties of generating function (similar to Laplace transforms)
leads to algebraic singularities of $R$ and $\hR$ around $x=1$~\cite{Flajolet}:
\begin{eqnarray}
R(1-s) &\sim& 1 - \phi_{\rm p} - A \ \Gamma(1-a) s^a \ , \\
\hR(1-s) &\sim& 1 - \phi_{\rm p}^{k-1} 
- (k-1) \phi_{\rm p}^{k-2} A \ \Gamma(1-a) s^a
- \frac{(k-1)(k-2)}{2} \phi_{\rm p}^{k-3} A^2 \ \Gamma(1-2a) s^{2a}
\end{eqnarray}
where the equivalent notation hold in the $s\to 0$
limit, and $\Gamma$ is Euler's special function. Inserting these expressions
in Eq.~(\ref{eq_app_R}), one can expand in powers of $s$ and identify
the terms of order $s^0$, $s^a$ and $s^{2 a}$ on both sides of the
equation. The first two orders compensate because of, respectively, the
relation on the order parameter (\ref{eq_app_order}) and its derivative
(\ref{eq_app_derivative}). The order $s^{2a}$ fixes the exponent $a$ under the
form (\ref{eq_MCT}), with $\lambda$ given by (\ref{eq_lambda_xorsat}).

We now consider the limit $\alpha \to \alpha_{\rm p}$ and denote 
$\delta = \alpha_{\rm p} - \alpha $ the (vanishing) distance to the 
transition. There are two scaling regimes
to be distinguished; the first governs the behavior of $Q_n$ in the
neighborhood of the plateau. Suppose this regime is reached on a scale
$n_{\rm i}(\delta)$ diverging with $\delta$ and described by the following
scaling function:
\begin{equation}
\epsilon(t) = \lim_{\delta \to 0} \delta^{-1/2} 
[Q_{n=t n_{\rm i}(\delta)} - \phi_{\rm p} ] \ .
\label{eq_scaling1}
\end{equation}
Expanding Eqs.~(\ref{eq_app_hQ},\ref{eq_app_R}) order by order in $\delta$,
one finds similarly (see~\cite{MoSe} for details) that the two first orders
are satisfied thanks to relations (\ref{eq_app_order},\ref{eq_app_derivative}),
while the third leads to an integro-differential equation for the scaling
function $\epsilon(t)$. The important feature of $\epsilon(t)$ is
its behavior in the small and large $t$ limits (entrance and exit from
the plateau):
\begin{equation}
\epsilon(t) \underset{t \to 0}{\sim} t^{-a} \ , \qquad  \qquad  
\epsilon(t) \underset{t \to \infty}{\sim} t^{b} \ ,
\label{eq_behav_eps}
\end{equation}
where $a$ is the same exponent as before, and $b$ the dual one 
(cf. Eq.~(\ref{eq_MCT})). In fact the small $t$ behavior of $\epsilon$
allows to fix the still undetermined scale $n_{\rm i}(\delta)$: for a large, 
yet independent of $\delta$, value of $n$, the study right at $\alpha_{\rm p}$
lead to $Q_n - \phi_{\rm p} \sim n^{-a}$. For consistency we must have
$n^{-a} \sim \delta^{1/2} (n/n_{\rm i}(\delta))^{-a}$, which implies 
$n_{\rm i}(\delta) \sim \delta^{-1/2a}$.

The second scaling regime describes the decay of $Q_n$ from its plateau value
down to zero, i.e. the distribution of the almost-frozen rearrangements whose
size is diverging as $\alpha$ reaches $\alpha_{\rm p}$. Suppose again
the existence of another scale $n_{\rm f}(\delta)$ for this to happen,
and of the scaling function
\begin{equation}
Q(t) = \lim_{\delta \to 0} Q_{n = t n_{\rm f}(\delta)} \ .
\label{eq_scaling2}
\end{equation}
Plugging this ansatz in Eqs.~(\ref{eq_app_hQ},\ref{eq_app_R}) one
obtains another equation for $Q(t)$, which implies in particular
$Q(t)- \phi_{\rm p} \sim t^b$ at small $t$. Matching the small $t$ behavior
of $Q(t)$ with the large $t$ limit of the previous scaling function 
$\epsilon(t)$, one finds that $n_{\rm f}(\delta) \sim \delta^{-\nu}$,
with $\nu = (1/2a) + (1/2b)$, as announced in the main part of the text.

\subsection{SAT}
\label{sec_app_sat}

The same steps, with some technical adaptations, can be followed in the 
case of SAT.
Let us first define the integrated distributions and the generating functions
for each value of the conditioning field:
\begin{equation}
Q_n(h) = \sum_{n'\ge n} q_{n'}(h) \ , \qquad
\hQ_n(u) = \sum_{n'\ge n} \hq_{n'}(u) \ , \qquad
R(h,x) = \sum_n q_n(h) x^n \ , \qquad 
\hR(u,x) = \sum_n \hq_n(u) x^n
\end{equation}
We rewrite Eqs.~(\ref{eq_hq_sat},\ref{eq_q_sat},\ref{eq_sat_final}) 
as
\begin{eqnarray}
\hQ_n(u) \hPP(u) &=& \int \prod_{i=1}^{k-1} d\P(h_i) 
\ \delta(u-f(h_1,\dots,h_{k-1}))  
\prod_{i=1}^{k-1} \frac{1-\tanh h_i}{2} Q_{n}(h_i) 
\qquad {\rm for} \ n \ge 1 \ , 
\label{eq_minsize_Q} \\
R(h,x) \P(h) &=& x 
\sum_{l_+,l_-=0}^\infty p_{l_+,l_-}
\int \prod_{i=1}^{l_+} d\hPP(u_i^+) \prod_{i=1}^{l_-} d\hPP(u_i^-) \ 
\delta\left(h -\sum_{i=1}^{l_+} u_i^+ + \sum_{i=1}^{l_-} u_i^- \right)
\prod_{i=1}^{l_-} \hR(u_i^-,x) \ ,
\label{eq_minsize_R} \\
Q_n &=& \int d\P(h) (1 - \tanh h) Q_n(h) \ .
\end{eqnarray}
Recall that the functional order parameters $\phi(h)=\lim Q_n(h) = 1-R(h,x=1)$
and $\hphi(u)$ are solutions of the 
equations~(\ref{eq_param_sat_htou},\ref{eq_param_sat_utoh}); we denote
$\phi_{\rm p}(h)$ and $\hphi_{\rm p}(u)$ their values at the threshold
$\alpha_{\rm p}$ for the appearance of a non-trivial solution, and
$\phi_{\rm p}= \lim Q_n = \int d\P(h) (1-\tanh h) \phi_{\rm p}(h)$.

For our purposes it will be sufficient to work with the simplified versions of
Eqs.~(\ref{eq_minsize_Q},\ref{eq_minsize_R}) obtained by integration over the
fields:
\begin{eqnarray}
\int d\hPP(u) \hQ_n(u) &=& 
\left(\int d\P(h) \frac{1-\tanh h}{2} Q_n(h) \right)^{k-1} 
= \frac{1}{2^{k-1}} Q_n^{k-1} \ , \label{eq_minsize_Qint} \\
\int d\P(h) R(h,x) &=& x \exp \left[-\frac{\alpha k}{2} 
+ \frac{\alpha k}{2}\int d\hPP(u) \hR(u,x) \right] \ . \label{eq_minsize_Rint}
\end{eqnarray}

Consider now the behavior of these quantities right at the transition 
$\alpha_{\rm p}$.
The simplest hypothesis is to assume the existence of a single exponent $a$ 
describing the decay of the integrated distributions 
$Q_n(h)$, $\hQ_n(u)$, towards their limit (as $n \to \infty$) $\phi(h)$, 
$\hphi(u)$, independently of $h,u$. This hypothesis is customary in the 
formally analog mode coupling theory of liquids~\cite{MCT}, where the role 
of the conditioning field is held by a wave vector.
We thus make the ansatz
$Q_n(h) \sim \phi(h) + A(h) n^{-a}$ with $A(h)$ a positive function.
Expanding Eq.~(\ref{eq_minsize_Qint}), this leads to
\begin{eqnarray}
\int d\hPP(u) \hQ_n(u) \sim \left( \frac{\phi_{\rm p}}{2} \right)^{k-1}
&+& \frac{k-1}{2^{k-1}} \phi_{\rm p}^{k-2} 
\left(\int d\P(h) (1-\tanh h) A(h) \right) n^{-a}
\\
&+& \frac{(k-1)(k-2)}{2^k} \phi_{\rm p}^{k-3}
\left(\int d\P(h) (1-\tanh h) A(h) \right)^2 n^{-2a} \ .
\end{eqnarray}
These algebraic decays at large $n$ translate into singularities in the
generating function around $x=1$,
\begin{eqnarray}
\int d\P(h) R(h,1-s) \sim 1 &-& \int d\P(h) \phi_{\rm p}(h) 
-  \left(\int d\P(h) A(h)\right) \Gamma(1-a) s^a \ ,\\
\int d\hPP(u) \hR(u,1-s) \sim 1 &-& 
\left( \frac{\phi_{\rm p}}{2} \right)^{k-1}
- \frac{k-1}{2^{k-1}} \phi_{\rm p}^{k-2}
\left(\int d\P(h) (1-\tanh h) A(h) \right) \Gamma(1-a) s^a
\\
&-& \frac{(k-1)(k-2)}{2^k} \phi_{\rm p}^{k-3}
\left(\int d\P(h) (1-\tanh h) A(h) \right)^2 \Gamma(1-2a) s^{2a} \ .
\end{eqnarray}
Finally these expansions are inserted in (\ref{eq_minsize_Rint});
collecting the terms of order $s^0$, $s^a$, $s^{2a}$ 
yields the following three equations :
\begin{eqnarray}
\int d\P(h) \phi_{\rm p}(h) &=& 
1- \exp \left[ -\frac{\alpha_{\rm p} k}{2^k} \phi_{\rm p}^{k-1}\right] \ ,\\
\int d\P(h) A(h) &=& \frac{\alpha_{\rm p} k(k-1)}{2^k} 
\phi_{\rm p}^{k-2} 
\exp\left[ -\frac{\alpha_{\rm p} k}{2^k} \phi_{\rm p}^{k-1}\right]
\left( \int d\P(h) (1 - \tanh h) A(h) \right) \ ,
\label{eq_minsize_ordersa} \\
\frac{\Gamma(1-a)^2}{\Gamma(1-2a)} &=& \lambda 
= \frac{2^k(k-2)}{\alpha_{\rm p} k (k-1)\phi_{\rm p}^{k-1} } \ . 
\label{eq_minsize_orders2a}
\end{eqnarray}

The first is a direct consequence of the equations 
(\ref{eq_param_sat_htou},\ref{eq_param_sat_utoh}) on the order parameter,
and can also be derived from (\ref{eq_minsize_Qint},\ref{eq_minsize_Rint}), 
setting $x=1$ in the latter.

The second is a functional analog of (\ref{eq_app_derivative}) 
and deserves a short explanation. The order parameter $\phi(h)$ is defined 
as the solution of a fixed-point functional equation of the type 
$\phi_\alpha=V[\phi_\alpha,\alpha]$, where we keep implicit the functional 
character of $\phi$ but emphasize the dependence on the control parameter 
$\alpha$. The relevant non-trivial solution of this equation which exists
for $\alpha \ge \alpha_{\rm p}$ disappears at $\alpha_{\rm p}$: this is
a bifurcation point in the vocabulary of discrete dynamical systems. A powerful
tool in this context is the implicit function theorem: if for some value 
$\alpha_0$ there is a solution $\phi_{\alpha_0}$ and if the differential
of $V$ with respect to $\phi$ in $(\phi_{\alpha_0},\alpha_0)$ has no 
eigenvector
of eigenvalue 1, then the solution $\phi_\alpha$ can be continuously followed
in a neighborhood of $\alpha_0$. At the bifurcation point $\alpha_{\rm p}$ the
hypothesis of the theorem must be violated. Linearizing 
Eqs.~(\ref{eq_param_sat_htou},\ref{eq_param_sat_utoh}), 
the reader will easily verify that an eigenvector
of eigenvalue 1 of the differential satisfies Eq.~(\ref{eq_minsize_ordersa}).
We can thus assume $A(h)$ to be in this eigenspace for the second condition
to be verified\footnote{this explanation is of course heuristic; the 
functional character of the fixed poind equation makes the invocation
of the implicit function theorem rather fuzzy.}.
Note that for a real order parameter equation $\phi=V(\phi,\alpha)$, this 
condition is nothing but the equality of the derivatives $1=\partial_\phi V$ 
at a transition, as used for instance in (\ref{eq_app_derivative}).

The third equation fixes the exponent $a$ and gives the value of the
exponent $\lambda$, as was claimed in the main part of the text 
(cf. Eqs.~(\ref{eq_MCT}) and (\ref{eq_lambda_sat})).

\begin{figure}
\includegraphics[width=8cm]{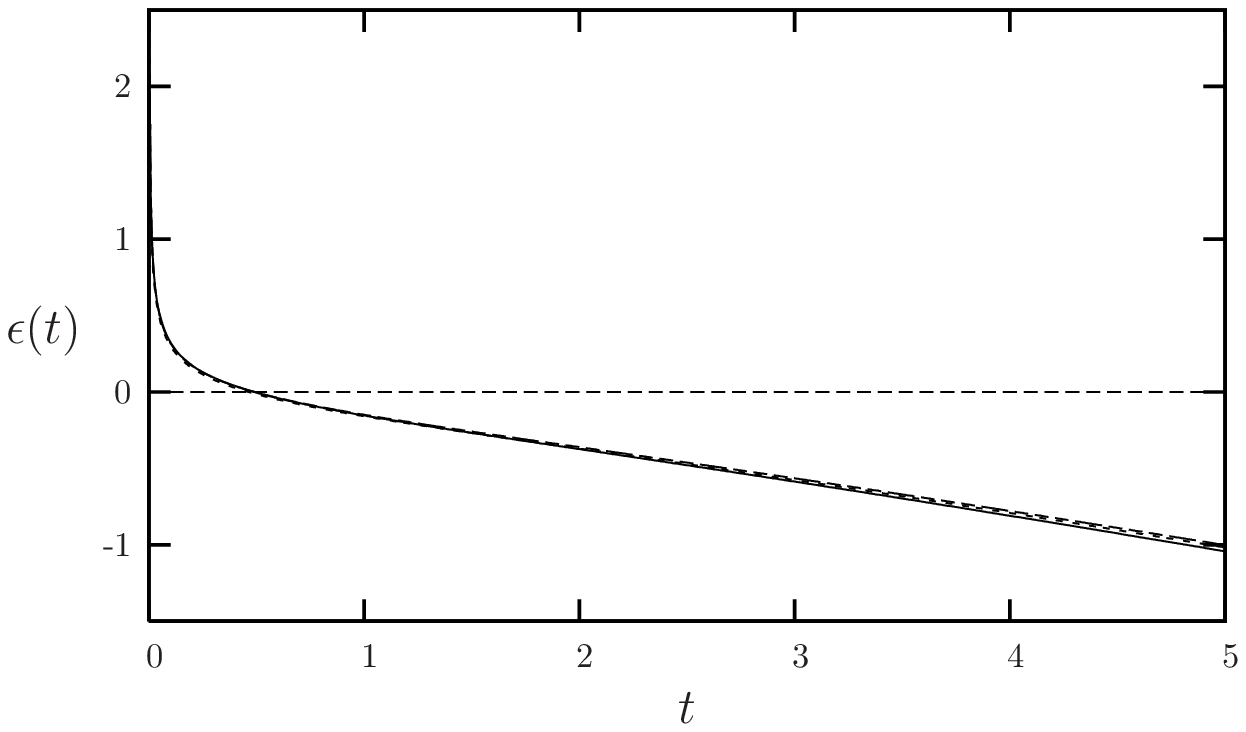} \hspace{1cm}
\includegraphics[width=8cm]{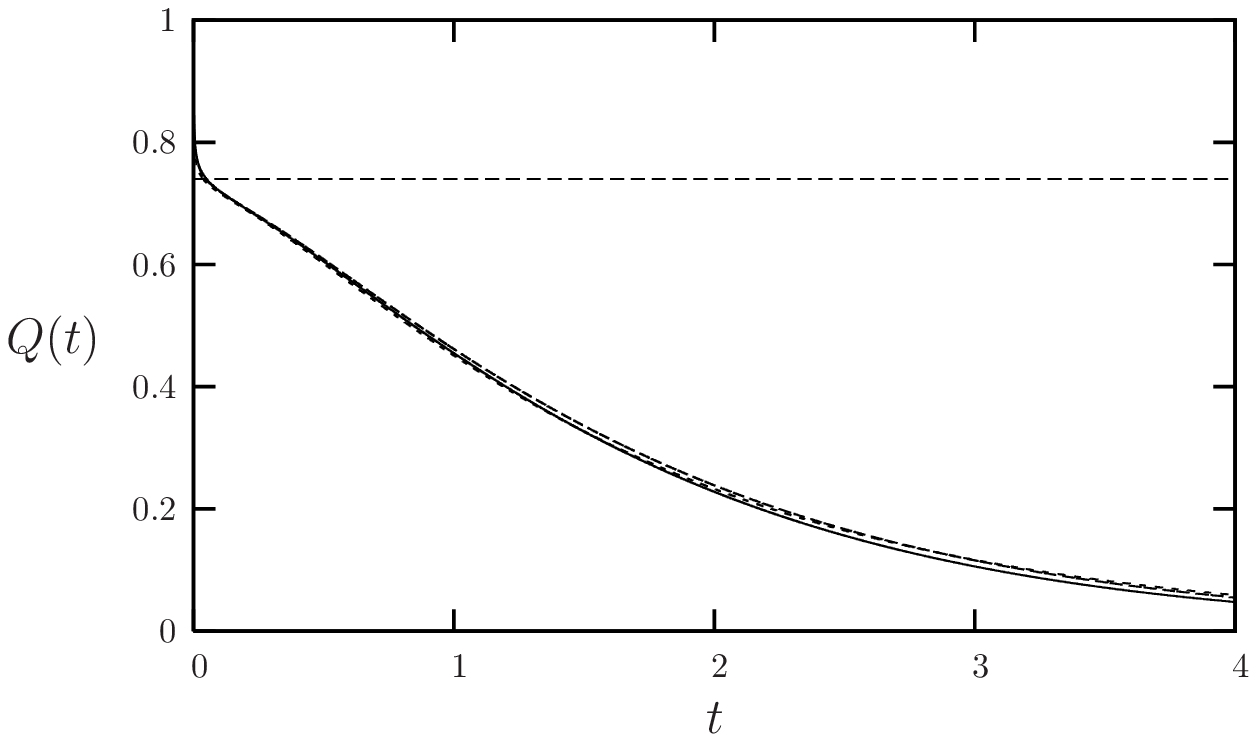}
\caption{The scaling functions of the average m.s.r.d. for the random tree 
ensemble of 3-SAT. The almost superimposed curves correspond to 
$\alpha = 4.39$, $4.392$, $4.396$.
Left: intermediate scale $t = n (\alpha_{\rm p}-\alpha)^{1/2a}$, 
see Eq.~(\ref{eq_scaling1}).
Right: final scale $t = n (\alpha_{\rm p}-\alpha)^\nu$, 
cf. Eq.~(\ref{eq_scaling2}), the dashed horizontal line indicates the
order parameter $\phi_{\rm p}$.
Numerical values of the exponents can be found in Table \ref{tab}.}
\label{fig_rescaled}
\end{figure}

The study of the intermediate and final scaling regimes can be done
following the lines sketched above on the XORSAT example;
for instance
the behavior around the plateau is described, for all values of the
cavity fields, by a single scaling function, generalizing 
Eq.~(\ref{eq_scaling1}) to
\begin{equation}
Q_{n=t n_{\rm i}(\delta)}(h) \sim \phi_{\rm p}(h) 
+ \delta^{1/2} A(h) \epsilon(t) \ .
\end{equation}
Provided $A(h)$ is chosen in such a way that Eq.~(\ref{eq_minsize_ordersa})
is verified, $\epsilon(t)$ obeys the same kind of integro-differential 
equation as the scaling function of the XORSAT problem, and in particular
its behavior at small and large $t$ is identical (see 
Eq.~(\ref{eq_behav_eps})). We thus reach exactly the same conclusions
on the behavior of $n_{\rm i}(\delta)$ and $n_{\rm f}(\delta)$.
This is confirmed in Fig.~\ref{fig_rescaled},
which shows, in the two regimes, a good collapse of
numerically determined distributions $Q_n$ for three values of $\alpha$
approaching $\alpha_{\rm p}$.

\subsection{Asymptotics at large $k,q$}
\label{sec_app_largek}

We justify now the statements made in the main part of the text on
the large $k,q$ behavior of the freezing thresholds. This analysis
is simple in the case of XORSAT: from 
(\ref{eq_app_order},\ref{eq_app_derivative}) one obtains
a closed equation on the order parameter at the transition,
\begin{equation}
\frac{1}{k-1} = - 
\frac{(1-\phi_{\rm p}(k)) \ln(1-\phi_{\rm p}(k))}{\phi_{\rm p}(k) } \ ,
\end{equation}
which can be inverted to obtain an asymptotic expansion of 
$\phi_{\rm p}(k)$. Reinserting it in Eq.~(\ref{eq_app_derivative}) yields
\begin{equation}
\alpha_{\rm p}(k) = \frac{1}{k} \left( \ln k + \ln \ln k + 1 + 
O\left( \frac{\ln\ln k}{\ln k} \right) \right) \ .
\end{equation}
The formal correspondence with the COL problem (see Eq.~(\ref{eq_param_col}))
leads immediately to the left hand side of (\ref{eq_col_largeq}).

The distributions of fields $\P(h)$, $\hP(u)$ for random SAT formulas can
be shown from (\ref{eq_sat_RS}) to concentrate in the large $k$ limit around, 
respectively, $0$ and $2^{-k}$. The equations (\ref{eq_hq_sat},\ref{eq_q_sat})
on $q_n(h)$, $\hq_n(u)$ can thus be simplified at the leading order in $k$
by retaining only these deterministic values of the conditioning fields. 
A simple transformation then shows that the distribution $q_n(h=0)$ collapses 
onto the solution of the XORSAT equations 
(\ref{eq_hq_XORSAT_Linfty},\ref{eq_q_XORSAT_Linfty}), provided the 
connectivity $\alpha$ is divided by a factor of $2^k$. This leads to the 
asymptotic behavior of the freezing threshold stated in 
Eq.~(\ref{eq_sat_largek}), and to the equivalence at large
$k$ of the exponents $a$, $b$, $\nu$ in the SAT and XORSAT problems.
A systematic expansion in powers of $2^{-k}$ of the deviations between
the two models could be set up from this starting point.

\section{Minimal size rearrangements at the 1RSB level}
\label{sec_app_1rsb}

\subsection{General case}

We consider in this appendix the computation proposed in
Sec.~\ref{sec_msr_1rsb}, namely the determination of the m.s.r.d.
for a finite tree factor graph whose initial configuration is drawn according
to the law $\mu^{(1)}$ (see Eq.~(\ref{eq_def_mu1})). To characterize
it we introduce on each directed edge of the factor graph a distribution of
cavity fields, denoted $P_{i \to a}(\eta)$ and $\hP_{a \to i}(\nu)$. They
obey the following set of equations,
\begin{eqnarray}
\hP_{a \to i}(\nu) &=& \frac{1}{Z(\{ P_{j \to a} \})}
\int \prod_{j\in \partial a \setminus i} dP_{j \to a}(\eta_{j \to a})
\ \delta( \nu -f(\{\eta_{j \to a} \}) )
\ z(\{\eta_{j \to a} \})^m \ ,
\\
P_{i \to a}(\eta) &=& \frac{1}{Z(\{ \hP_{b \to i} \})}
\int \prod_{b \in \partial i \setminus a}
d\hP_{b \to i}(\nu_{b \to i}) \ \delta(\eta - g(\{ \nu_{b \to i} \} ))
\ z(\{ \nu_{b \to i} \} )^m \ ,
\end{eqnarray}
where the functions $f$, $g$ and $z$ are the ones defined in
(\ref{eq_def_f},\ref{eq_def_g}) for the corresponding edges. The
boundary condition is given by $P_{i \to a}(\eta) = P_{{\rm ext},i}(\eta)$ if 
$i\in B$, otherwise $P_{i \to a}(\eta) = \delta(\eta-\oeta)$. The marginals
of $\mu^{(1)}$ can be obtained from these distributions, for instance for
a single variable one obtains
\begin{equation}
\mu^{(1)}(\s_i) = \int dP_i(\eta) \ \eta(\s_i) \ , \qquad
P_i(\eta) = \frac{1}{Z(\{ \hP_{a \to i} \})}
\int \prod_{a \in \partial i}
d\hP_{a \to i}(\nu_{a \to i}) \ \delta(\eta - g(\{ \nu_{a \to i} \} ))
\ z(\{ \nu_{a \to i} \} )^m
 \ .
\end{equation}
We also have to introduce distributions of the size messages, 
$q_{\vec{n}}^{(i \to a,\s_i)}(\eta)$ and
$\hq_{\vec n}^{(a\to i,\s_i)}(\nu)$, which corresponds to the weighted
averages of the distributions in a single $\mu^{(0)}$.
From (\ref{eq_broadcast2},\ref{eq_q_itoa_gene}) one obtains
\begin{multline}
\hP_{a \to i}(\nu)\ \hq_{\vec n}^{(a\to i,\s_i)}(\nu) = 
\frac{1}{Z(\{ P_{j \to a} \})}
\int \prod_{j\in \partial a \setminus i} dP_{j \to a}(\eta_{j \to a})
\ \delta( \nu -f(\{\eta_{j \to a} \}) ) \ z(\{\eta_{j \to a} \})^m \\
\sum_{\us_{a \setminus i}} 
\mu(\us_{a \setminus i}  | \s_i, \{ \eta_{j \to a} \} )
\prod_{j \in \partial a \setminus i} 
\sum_{\vec{n}_{j \to a}}
q^{(j \to a,\s_j)}_{\vec{n}_{j \to a}} (\eta_{j \to a} )
\ \delta_{\vec{n},\tf(\{\vec{n}_{j \to a}\}) }
\end{multline}
and
\begin{multline}
P_{i \to a}(\eta) \ q_{\vec{n}}^{(i \to a,\s_i)}(\eta)
= \frac{1}{Z(\{ \hP_{b \to i} \})}
\int \prod_{b \in \partial i \setminus a}
d\hP_{b \to i}(\nu_{b \to i}) \ \delta(\eta - g(\{ \nu_{b \to i} \} )) \
z(\{ \nu_{b \to i} \} )^m \\ \prod_{b \in \partial i \setminus a} 
\sum_{\vec{n}_{b \to i}}
\hq^{(b \to i,\s_i)}_{\vec{n}_{b \to i}}(\nu_{b \to i})
\ \delta_{\vec{n},
\tg_{\s_i}(\{ \vec{n}_{b \to i} \} )} \ ,
\end{multline}
with the boundary condition at the leaves
$q_{\vec{n}}^{(i \to a,\s)}(\eta) = \delta_{\vec{n},\vec{o}(\s)}$.
Finally the m.s.r.d. with respect to $\mu^{(1)}$ for a variable $i$
reads
\begin{equation}
q_n^{(i)} = \int dP_i(\eta) \sum_{\s_i} \eta(\s_i) \sum_{\vec{n}}
q_{\vec{n}}^{(i,\s_i)}(\eta)
\ \delta_{n,\underset{\t_i \neq \s_i}{\min}[\vec{n}]_{\t_i}} \ ,
\end{equation}
with
\begin{multline}
P_i(\eta) \ q_{\vec{n}}^{(i,\s_i)}(\eta)
= \frac{1}{Z(\{ \hP_{a \to i} \})}
\int \prod_{a \in \partial i}
d\hP_{a \to i}(\nu_{a \to i}) \ \delta(\eta - g(\{ \nu_{a \to i} \} )) \
z(\{ \nu_{a \to i} \} )^m \\ \prod_{a \in \partial i} 
\sum_{\vec{n}_{a \to i}}
\hq^{(a \to i,\s_i)}_{\vec{n}_{a \to i}}(\nu_{a \to i})
\ \delta_{\vec{n},
\tg_{\s_i}(\{ \vec{n}_{a \to i} \} )} \ .
\end{multline}
Note that this computation reduces to the one of Sec.~\ref{sec_gene_distri}
either when the distribution of cavity fields are concentrated on a single
value or when $m=1$, defining in the latter case
\begin{equation}
\eta_{i \to a} = \int dP_{i \to a}(\eta) \ \eta \ , \qquad
q_{\vec{n}}^{(i \to a,\s_i)} = \frac{\int dP_{i \to a}(\eta) \ \eta(\s_i) \ 
q_{\vec{n}}^{(i \to a,\s_i)}(\eta)}{\eta_{i \to a}(\s_i)} \ ,
\label{eq_reduc_m1}
\end{equation}
and similarly $\nu_{a \to i}$ and $\hq_{\vec{n}}^{(a \to i,\s_i)}$. 
For a generic value of $m$ one proceeds with the computation of the
average m.s.r.d. for a random tree; the only modification with respects to
Sec.~\ref{sec_gene_ave} is a replacement of the distribution of external
fields $\P(\eta)$ by a distribution of distribution of fields, $\P(P)$.
One has thus to define $q_{\eta,\vec{n}}^{(\s,L)}(P)$, the average of
the joint law $P_i(\eta) q_{\vec{n}}^{(i,\s_i)}(\eta)$ on $\eta$ and 
$\vec{n}$ for the root of $\T_L$, conditioned
on the event $P=P_i$, and similarly $\hq_{\nu,\vec{n}}^{(\s,L)}(\hP)$ for
$\hT_L$. These quantities can be obtained by recursions on $L$ through
equations formally similar to (\ref{eq_hq_gene},\ref{eq_q_gene}), which could
in principle be solved numerically using a population of
population of elements $(\eta,\vec{n}^{(1)},\dots,\vec{n}^{(q)})$. 
We shall give
the explicit form of these equations in the two particular cases of
SAT and COL in the following two subsections.

\subsection{SAT}

For random SAT instances the stationarity conditions for the distribution
of distribution of fields $\P(P)$, $\hPP(\hP)$ can be written in their
distributional form as
\begin{equation}
\hP \eqd F(P_1,\dots,P_{k-1}) \ , \qquad
P \eqd G(\hP_1^+,\dots,\hP_{l_+}^+,\hP_1^-,\dots,\hP_{l_-}^-) \ .
\label{eq_1RSB_distrib}
\end{equation}
The functionals $F$ and $G$ are defined by
\begin{eqnarray}
\hP(u) &=& \frac{1}{Z(\{ P_i \})} \int \prod_{i=1}^{k-1}dP_i(h_i)
\ \delta(u - f(h_1,\dots,h_{k-1})) \ z(h_1,\dots,h_{k-1})^m \ ,
\label{eq_1RSB_hP} \\
P(h) &=& \frac{1}{Z(\{ \hP_i^\pm \})}
\int \prod_{i=1}^{l_+}d\hP_i^+(u_i^+) \prod_{i=1}^{l_-}d\hP_i^-(u_i^-)
\ \delta\left( h - \sum_{i=1}^{l_+} u_i^+ + \sum_{i=1}^{l_-} u_i^-  \right) 
z(u_1^+,\dots,u_{l_+}^+,u_1^-,\dots,u_{l_-}^- )^m \ ,
\label{eq_1RSB_P}
\end{eqnarray}
where
\begin{eqnarray}
&&z(h_1,\dots,h_{k-1}) = 2 - \prod_{i=1}^{k-1} \frac{1 - \tanh h_i}{2} \ ,\\
&&z(u_1^+,\dots,u_{l_+}^+,u_1^-,\dots,u_{l_-}^- ) = 
\prod_{i=1}^{l_+} \frac{1+\tanh u_i^+}{2} 
\prod_{i=1}^{l_-} \frac{1-\tanh u_i^-}{2} + 
\prod_{i=1}^{l_+} \frac{1-\tanh u_i^+}{2} 
\prod_{i=1}^{l_-} \frac{1+\tanh u_i^-}{2} \ .
\end{eqnarray}
The conditional average of the joint law of cavity field and sizes obey
the two following equations,
\begin{multline}
\hq_{u,n}^{(L)}(\hP) \hPP(\hP) = \int \prod_{i=1}^{k-1} d\P(P_i) 
\ \delta(P-F(\{P_i\} ) ) \frac{1}{Z(\{P_i\} )}
\int \prod_{i=1}^{k-1} dh_i \ \delta(u -f(h_1,\dots,h_{k-1})) \ 
z(h_1,\dots,h_{k-1})^m \\
\sum_{n_1,\dots,n_{k-1}} 
\prod_{i=1}^{k-1} q_{h_i,n_i}^{(L)}(P_i)
\left[ 
\left(1-\prod_{i=1}^{k-1} \frac{1 -\tanh h_i}{2} \right) \delta_{n,0}
+ \left( \prod_{i=1}^{k-1} \frac{1-\tanh h_i}{2} \right)
\delta_{n,\min[n_1,\dots,n_{k-1}]}
\right] \ ,
\label{eq_1RSB_hqn}
\end{multline}
\begin{multline}
q_{h,n}^{(L+1)}(P) \P(P) = \sum_{l_+,l_-} p_{l_+,l_-} \int 
\prod_{i=1}^{l_+} d\hPP(\hP_i^+) \prod_{i=1}^{l_-} d\hPP(\hP_i^-)
\ \delta(P-G(\{ \hP_i^\pm\}) 
\frac{1}{Z(\{ \hP_i^\pm\})} \\
\int \prod_{i=1}^{l_+} du_i^+ \prod_{i=1}^{l_-} du_i^-
 \ \delta\left( h - \sum_{i=1}^{l_+} u_i^+ + \sum_{i=1}^{l_-} u_i^-  \right) 
z(\{u_i^\pm\})^m 
\prod_{i=1}^{l_+} \hP_i^+(u_i^+)
\sum_{n_1,\dots,n_{l_-}} \prod_{i=1}^{l_-} \hq_{u_i^-,n_i}^{(L)}(\hP_i^-)
\delta_{n,1+n_1+\dots+n_{l_-}} \ .
\label{eq_1RSB_qn}
\end{multline}
These equations conserve the conditions $\sum_n q_{h,n}^{(L)}(P)=P(h)$ which
follow from the definition of $q_{h,n}^{(L)}(P)$. Finally the average m.s.r.d.
for the root of $\T_L$ reads
\begin{equation}
q_n^{(L)} = \int d\P(P) \int dh \ (1 - \tanh h) \ q_{h,n}^{(L)}(P) \ .
\label{eq_1RSB_qn_final}
\end{equation}
As a consistency check one can reduce these equations to the ones developed 
in the main part of the text 
(cf. (\ref{eq_hq_sat},\ref{eq_q_sat},\ref{eq_sat_final})) 
when $m=1$, using the identity (\ref{eq_reduc_m1}).

Let us come back on the 1RSB equations 
(\ref{eq_1RSB_distrib},\ref{eq_1RSB_hP},\ref{eq_1RSB_P}). It is possible
for the distributions $\hP(u)$ in the support of $\hPP$ to acquire a peak
on the hard field value $u=+\infty$, of intensity denoted $\hphi(\hP)$. This
corresponds to a field forcing the variable node to satisfy the constraint node
emitting the message. Similarly we call $\phi(P)$ the intensity of the peak
in $h=-\infty$, signaling a clause that the emitting variable is forced to
unsatisfy it. These intensities are found from 
(\ref{eq_1RSB_distrib},\ref{eq_1RSB_hP},\ref{eq_1RSB_P}) to obey
\begin{eqnarray}
\hphi(\hP) \hPP(\hP) &=& \int \prod_{i=1}^{k-1} d\P(P_i) 
\ \delta(P-F(\{P_i\} ) ) \frac{1}{Z(\{P_i\} )} \prod_{i=1}^{k-1} \phi(P_i) 
\ , \label{eq_1RSB_hphi} \\
\phi(P) \P(P) &=& \sum_{l_+,l_-} p_{l_+,l_-} \int 
\prod_{i=1}^{l_+} d\hPP(\hP_i^+) \prod_{i=1}^{l_-} d\hPP(\hP_i^-)
\ \delta(P-G(\{ \hP_i^\pm\}) 
\frac{1}{Z(\{ \hP_i^\pm\})} \label{eq_1RSB_phi} \\ && \hspace{-5mm} 
\prod_{i=1}^{l_+} \int d\hP_i^+(u) \left(\frac{1-\tanh u}{2}\right)^m
\prod_{i=1}^{l_-} \int d\hP_i^-(u) \left(\frac{1+\tanh u}{2}\right)^m
\left[ 1 - \prod_{i=1}^{l_-} \left( 1 - 
\frac{\hphi(\hP_i^-)}{\int d\hP_i^-(u) \left(\frac{1+\tanh u}{2}\right)^m}
\right) \right] \ . \nonumber
\end{eqnarray}
A randomly chosen variable will receive a forcing hard field in a randomly
chosen pure state with probability
\begin{equation}
\phi = 2 \int d\P(P) \phi(P) \ ,
\end{equation}
where the factor $2$ comes from the symmetry between positive and negative
literals. $\phi$ is also the order parameter of the freezing transition; the
equations (\ref{eq_1RSB_hqn},\ref{eq_1RSB_qn}), in the $L \to \infty$ limit,
admit a solution where $\phi(P)$ (resp. $\hphi(\hP)$) is the intensity
of a Dirac peak on $(h,n)=(-\infty,\infty)$ (resp. $(u,n)=(+\infty,\infty)$).
The fraction of diverging rearrangements in (\ref{eq_1RSB_qn_final}) is then
seen to be equal to $\phi$.

In order to discuss the critical behavior of the m.s.r.d. it is convenient
to derive an integrated version of Eqs.~(\ref{eq_1RSB_hqn},\ref{eq_1RSB_qn}),
\begin{equation}
\int d\hPP(\hP) 
\frac{\int du \; \hQ_{u,n}(\hP) (1+\tanh u)^m}{\int d\hP(u) (1+\tanh u)^m} =
\left(\int d\P(P) \int dh \frac{1-\tanh h}{2} Q_{h,n}(P)\right)^{k-1} 
= \frac{1}{2^{k-1}} Q_n^{k-1} \ ,
\label{eq_1RSB_integ1}
\end{equation}
\begin{equation}
\int d\P(P) 
\frac{\int dh \; R_{h,x}(P) (1-\tanh h)^m}{\int dP(h) (1-\tanh h)^m}
= x \exp \left[ -\frac{\alpha k}{2} + \frac{\alpha k}{2} \int d\hPP(\hP) 
\frac{\int du \; \hR_{u,x}(\hP) (1+\tanh u)^m}{\int d\hP(u) (1+\tanh u)^m}
\right] \ ,
\label{eq_1RSB_integ2}
\end{equation}
where the former is valid for $n \ge 1$ and following our conventions we
defined
\begin{equation}
Q_{h,n}(P) = \sum_{n' \ge n} q_{h,n'}(P) \ , \qquad
R_{h,x}(P) = \sum_n x^n q_{h,n}(P) \ .
\end{equation}
Let us call $\alpha_{\rm p}^{(1)}$ the threshold value for the appearance
of a non-trivial solution to (\ref{eq_1RSB_hphi},\ref{eq_1RSB_phi}), and
$\phi_{\rm p}^{(1)}$ the corresponding order parameter. We want to determine
the critical behavior of $q_n$ in the neighborhood of this threshold,
expecting to recover the phenomenology obtained in the $m=1$ case. For
simplicity we shall consider only the first critical regime at
$\alpha=\alpha_{\rm p}^{(1)}$, supposing an algebraic decay of $Q_n$ with
an exponent $a$ to its asymptotic value $\phi_{\rm p}^{(1)}$. More
precisely we make the ansatz $Q_{h,n}(P) = \delta(h+\infty) (\phi(P)
+ A(P) n^{-a}) + o(n^{-a})$, with $A(P)$ a positive function.
The computation proceeds as in Sec.~\ref{sec_app_sat}: one inserts this
ansatz in (\ref{eq_1RSB_integ1}) and expands to order $n^{-2a}$. The algebraic
decays translate into singularities around $x=1$ in the generating functions of
Eq.~(\ref{eq_1RSB_integ2}), matching the three leading orders one obtains
\begin{eqnarray}
\int d\P(P) \frac{\phi(P)}{\int dP(h) \left(\frac{1-\tanh h}{2}\right)^m} &=& 
1- \exp \left[ -\frac{\alpha_{\rm p}^{(1)} k}{2^k} (\phi_{\rm p}^{(1)})^{k-1}\right] \ ,\\
\int d\P(P) \frac{A(P)}{\int dP(h) \left(\frac{1-\tanh h}{2}\right)^m}
 &=& \frac{\alpha_{\rm p}^{(1)} k(k-1)}{2^k} 
(\phi_{\rm p}^{(1)})^{k-2} 
\exp
\left[ -\frac{\alpha_{\rm p}^{(1)} k}{2^k} (\phi_{\rm p}^{(1)})^{k-1}\right]
\int d\P(P) A(P)
\ , \\
\frac{\Gamma(1-a)^2}{\Gamma(1-2a)} &=& \lambda^{(1)} 
= \frac{2^k(k-2)}{\alpha_{\rm p}^{(1)} k (k-1)(\phi_{\rm p}^{(1)})^{k-1} } 
\ .
\end{eqnarray}
The first equality is a direct consequence of 
(\ref{eq_1RSB_hphi},\ref{eq_1RSB_phi}), the second is fulfilled by taking
$A(P)$ in the eigenspace of eigenvalue 1 of the differential of 
(\ref{eq_1RSB_hphi},\ref{eq_1RSB_phi}), while the third fixes the exponent
$a$. The computation of the parameter $\lambda$ at the RSB level thus leads to
the expression found in the RS approach (cf. Eq.~(\ref{eq_lambda_sat})), 
apart from the replacement of the critical connectivity and order parameter 
with their corresponding RSB values.

\subsection{COL}

The random $q$-COL model is described at the 1RSB level by a distribution
$\P(P)$ over (invariant under the color permutations) distributions $P(\eta)$ 
of fields (laws on $\X=\{1,\dots,q\}$).
$\P$ is solution of the distributional equation $P \eqd F(P_1,\dots,P_l)$,
where $l$ is a Poisson random variable of mean $c$ and $F$ is defined by
\begin{equation}
P(\eta) = \frac{1}{Z(P_1,\dots,P_l)} \int dP_i(\eta_i) \ 
\delta(\eta - f(\eta_1,\dots,\eta_l)) \ z(\eta_1,\dots,\eta_l)^m \ , \quad
f(\{\eta_i\})(\s) = \frac{1}{z(\{\eta_i\})}
\prod_{i=1}^l (1-\eta_i(\s)) \ .
\label{eq_1RSB_col}
\end{equation}
One can distinguish the hard fields
which constrain a variable to take a definite color and define
\begin{equation}
P(\eta) = \phi(P) \frac{1}{q} \sum_{\s=1}^q \delta(\eta - d_\s) +
(1-\phi(P)) \tP(\eta) \ , \qquad d_\s(\t)=\delta_{\s,\t} \ ,
\end{equation}
where $\tP$ is a normalized distributions with no intensity on the hard fields 
$d_\s$. The order parameter $\phi(P)$ is found from
(\ref{eq_1RSB_col}) to obey:
\begin{multline}
\phi(P) \P(P) = \sum_{l=0}^\infty p_l \int \prod_{i=1}^l d\P(P_i)
\ \delta(P-F(P_1,\dots,P_l))
\frac{1}{Z(P_1,\dots,P_l)} \\
\sum_{p=0}^{q-1} q \binom{q-1}{p}(-1)^p
\prod_{i=1}^l\left( 
\int dP_i(\eta) (1-\eta(\s))^m - \frac{p}{q} \phi(P_i) \right) \ .
\label{eq_1RSB_col_param}
\end{multline}

The average m.s.r.d. on random trees where the initial
configurations are drawn from the 1RSB measure $\mu^{(1)}$ reads
\begin{equation}
q_n^{(L)} = \int d\P(P) \int d\eta \ \sum_\s \eta(\s) \ q_{\eta,n}^{(\s,L)}(P)
\ ,
\end{equation}
where $q_{\eta,n}^{(\s,L)}(P)$ is the conditional average of the joint law
of size and fields messages. Note that all values of $\s$ contribute in the 
same way above, by the symmetry between colors. The equation governing
$q_{\eta,n}^{(\s,L)}(P)$ is
\begin{multline}
q_{\eta,n}^{(\s,L+1)}(P) \P(P) = \sum_{l=0}^\infty p_l 
\int \prod_{i=1}^l d\P(P_i) \ \delta(P-F(P_1,\dots,P_l)) 
\frac{1}{Z(P_1,\dots,P_l)} \\
\int \prod_{i=1}^l d\eta_i \ \delta(\eta-f(\eta_1,\dots,\eta_l)) \ 
z(\eta_1,\dots,\eta_l)^m 
\sum_{\substack{\s_1,\dots,\s_l \\ n_1,\dots,n_l}}
\prod_{i=1}^l \mu(\s_i|\s;\eta_i)\  q_{\eta_i,n_i}^{(\s_i,L)}(P_i) \
\I \left(n = 1+\min_{\t \neq \s} \sum_{i=1}^l \delta_{\t,\s_i} n_i\right) \ ,
\label{eq_1RSB_col_q}
\end{multline}
with
\begin{equation}
\mu(\s_i|\s;\eta_i) = \frac{\eta_i(\s_i)}{1-\eta_i(\s)} \I(\s_i \neq \s) \ .
\end{equation}

The order parameter $\phi = \int d\P(P) \phi(P)$ is again the height of the
plateau in the $L \to \infty$ limit of the integrated average m.s.r.d. $Q_n$.
One can indeed check that $q_{\eta,n}^{(\s)}(P)$ has a Dirac peak of intensity
$\phi(P)/q$ in $(\eta,n)=(d_\s,\infty)$.

The study of the critical behavior at the transition $c_{\rm p}^{(1)}$
corresponding to the appearance of hard fields in the 1RSB distributions
is similar to the SAT case. We first write an integrated version of 
(\ref{eq_1RSB_col_q}),
\begin{multline}
\int d\P(P) 
\frac{\int d\eta \ \eta(\s)^m \ q_{\eta,n}^{(\s)}(P)}
{\int dP(\eta)\ \eta(\s)^m }
= \sum_{l=0}^\infty \frac{e^{-c} c^l}{l!} \frac{1}{(q-1)^l} 
\sum_{\s_1,\dots,\s_l =2}^q \sum_{ n_1,\dots,n_l}
\I\left( n = 1+ \min_{\t=2\dots,q } \left[\sum_{i=1}^l \delta_{\t,\s_i} n_i 
\right] \right) \\
\prod_{i=1}^l \left[ (q-1) \int d\P(P) 
\frac{\int d\eta \ (1-\eta(1))^{m-1} \ \eta(\s_i) \ q_{\eta,n_i}^{(\s_i)}(P)}
{\int dP(\eta)\ (1-\eta(\s))^m } \right] \ ,
\end{multline}
which is independent on the value of $\s$.
The ansatz 
$Q_{\eta,n}^{(\s)}(P)=\delta(\eta - d_\s) (\phi(P) + A(P) n^{-a}) + o(n^{-a})$
is then inserted in this equation. The first two orders in an asymptotic
expansion at large $n$ in powers of $n^{-a}$ are satisfied thanks to
(\ref{eq_1RSB_col_param}) and by choosing $A(P)$ in the eigenspace
of eigenvalue 1 of its differential. The third order fixes the value of
the exponent $a$ through
\begin{equation}
\frac{\Gamma(1-a)^2}{\Gamma(1-2a)} = 
(q-2) \frac{1 - \tphi^{1/(q-1)}}{\tphi^{1/(q-1)}} \ ,
\qquad 
\tphi = \frac{1}{q} \int d\P(P) \frac{\phi(P)}{\int dP(\eta) \ \eta(\s)^m} \ ,
\end{equation}
which corresponds for $m=1$ to the expression found in 
Eq.~(\ref{eq_lambda_col}).


\end{document}